\documentclass[twoside,english,preprint,amsmath,nofootinbib]{revtex4}
\usepackage[T1]{fontenc}
\usepackage[latin9]{inputenc}
\usepackage{amsmath}
\usepackage{graphicx}
\usepackage{amssymb}
\usepackage{esint}

\makeatletter

\providecommand{\tabularnewline}{\\}

\@ifundefined{textcolor}{}
{%
 \definecolor{BLACK}{gray}{0}
 \definecolor{WHITE}{gray}{1}
 \definecolor{RED}{rgb}{1,0,0}
 \definecolor{GREEN}{rgb}{0,1,0}
 \definecolor{BLUE}{rgb}{0,0,1}
 \definecolor{CYAN}{cmyk}{1,0,0,0}
 \definecolor{MAGENTA}{cmyk}{0,1,0,0}
 \definecolor{YELLOW}{cmyk}{0,0,1,0}
 }

\@ifundefined{definecolor}{\@ifundefined{definecolor}
 {\usepackage{color}}{}
}{}\usepackage[dvipdfm,bookmarks=true]{hyperref}

\makeatother

\usepackage{babel}

\makeatother

\usepackage{babel}

\makeatother

\usepackage{babel}

\makeatother

\usepackage{babel}

\makeatother

\usepackage{babel}

\begin{document}

\title{{\LARGE Decaying Dark Matter in Supersymmetric SU(5) Models}}

\date{\today}

\author{Mingxing Luo}

\email{luo@zimp.zju.edu.cn}

\author{Liucheng Wang}

\email{liuchengwang@zimp.zju.edu.cn}

\author{Wei Wu}

\email{weiwu@zimp.zju.edu.cn}

\author{Guohuai Zhu}

\email{zhugh@zju.edu.cn}

\affiliation{Zhejiang Institute of Modern Physics, Department of Physics, Zhejiang
University, Hangzhou, Zhejiang 310027, P.R.China}
\begin{abstract}
Motivated by recent observations from Pamela, Fermi and H.E.S.S.,
we consider dark matter decays in the framework of supersymmetric
SU(5) grand unification theories. An SU(5) singlet $S$ is assumed
to be the main component of dark matters, which decays into visible
particles through dimension six operators suppressed by the grand
unification scale. Under certain conditions, $S$ decays dominantly
into a pair of sleptons with universal coupling for all generations.
Subsequently, electrons and positrons are produced from cascade decays
of these sleptons. These cascade decay chains smooth the $e^{+}+e^{-}$
spectrum, which permit naturally a good fit to the Fermi LAT data.
The observed positron fraction upturn by PAMELA can be reproduced
simultaneously. We have also calculated diffuse gamma-ray spectra
due to the $e^{\pm}$ excesses and compared them with the preliminary
Fermi LAT data from 0.1 GeV to 10 GeV in the region $0^{\circ}\leq l\leq360^{\circ},10^{\circ}\leq|b|\leq20^{\circ}$.
The photon spectrum of energy above 100 GeV, mainly from final state
radiations, may be checked in the near future.
\end{abstract}
\maketitle

\section{introduction}

Electron, proton, photon, neutrino and their antiparticles are
stable, at least on the cosmological time scale. Detection of these
particles from cosmic rays provides an interesting window to look
into the deep universe. Recently, the PAMELA experiment reported a
significant excess in the positron fraction $e^{+}/(e^{+}+e^{-})$
between $10$ GeV and $100$ GeV \cite{Adriani:2008zr}. On the other
hand, the measured antiproton to proton flux ratio appears to be
consistent with predictions \cite{Adriani:2008zq}. More recently,
the Fermi LAT collaboration observed a smooth $e^{+}+e^{-}$ spectrum
with high accuracy. It is found to be falling as $E^{-3.0}$ from
$20$ GeV to $1$ TeV \cite{Abdo:2009zk}, much harder than the
predictions of conventional models. The H.E.S.S. collaboration
measured the $e^{+}+e^{-}$ spectrum from $600$ GeV up to several TeV
\cite{Aharonian:2009ah}, which is consistent with the Fermi data in
overlapping regions and steepens at about $1$ TeV towards higher
energy.

These excesses of electrons and positrons could be due to unidentified
astrophysical sources, e.g., nearby pulsars or supernova remnants
\cite{Stawarz:2009ig,Piran:2009tx,Grasso:2009ma}. However, an explanation
via dark matter (DM) annihilation or decay is, arguably, a much more
interesting possibility, at least from the perspective of particle
physics. The electron and positron spectra alone, even with higher
precision and broader energy range, cannot decisively decide which
explanation is more plausible \cite{Malyshev:2009tw}. Hopefully,
the energy spectrum and the angular dependence of cosmic gamma rays
\cite{Bertone:2008xr,Zhang:2008tb,Bergstrom:2008ag,Ibarra:2009nw,Zhang:2009kp},
to be measured by the Fermi LAT in the near future, may provide a
more definite answer. For the DM interpretation, the mass of the DM
should be around several TeV, to provide the $e^{\pm}$ excesses from
$20$ GeV to $1$ TeV and steepen sharply above $1$ TeV. Furthermore,
traditional WIMP DM candidates usually produce extra antiprotons.
As Pamela does not observe any deviation on antiproton spectrum from
the anticipation, WIMP DMs are now disfavored as potential sources
of the observed cosmic-ray excesses. Still, there are plenty of freedoms
for both DM annihilation and decay to reproduce the experimental $e^{\pm}$
spectra reasonably \cite{Meade:2009iu,Bergstrom:2009fa}. For DM annihilation,
a large boost factor in the order of $10^{2}$ to $10^{3}$ is needed
for the theory to be consistent with the relic abundance measured
by the WMAP \cite{Dunkley:2008ie}. As the clumpiness property of
the DM distribution falls far short of such a large factor, one usually
resorts to nonperturbative Sommerfeld \cite{Hisano:2004ds,Hisano:2006nn,Cirelli:2007xd,ArkaniHamed:2008qn}
or Breit\textendash{}Wigner \cite{Feldman:2008xs,Ibe:2008ye,Guo:2009aj}
enhancement in model buildings. For DM decays, the lifetime should
typically be around the order of $10^{26}s$ to fit the $e^{\pm}$
data\cite{Meade:2009iu,Shirai:2009fq,Mardon:2009gw}, which is much
longer than the lifetime of the universe. Therefore the DM decay rates
will not affect the relic abundance appreciably.

The energetic $e^{\pm}$ flux produced from DM annihilations/decays
would inevitably emit gamma rays. These gamma rays depend on the DM
density as $\rho^{2}$ for annihilations and $\rho$ for decays. This
will lead to different angular dependence of the gamma ray spectrum,
which may be measurable in the near future to differentiate these
two scenarios. The gamma ray spectrum can also be used to differentiate
DM explanations from astrophysical ones.

In this paper, we will focused on DM decays. Notice that a lot of
suppression will be needed for a TeV scale particle to have a lifetime
$\sim10^{26}s$. If it decays via dimension four operators, tremendous
fine tunings will be needed. If it decays via dimension six operators,
it still needs to be suppressed by a scale $\sim10^{16}$ GeV, which
turns out to coincide with the grand unification theory (GUT) scale
\cite{Langacker:1991an,Amaldi:1991cn,Ellis:1990wk}. In the same spirit
of Refs. \cite{Arvanitaki:2008hq,Arvanitaki:2009yb}, we will take
a singlet as the dark matter candidate and provide a detailed analysis
in the frame of supersymmetric SU(5) GUT. To be consistent with the
Pamela antiproton measurement, squark masses are assumed to be heavier
than that of the SU(5) singlet $S$, so the $S$ decay would be quark
phobic. The $S$ then decays dominantly into slepton pairs with a
universal coupling for all generations. These sleptons decay quickly
into leptons and lightest supersymmetric particles (LSPs), if R-parity
is conserved. In this framework, we have obtained a reasonable fit
to all $e^{\pm}$ data from Pamela, Fermi and H.E.S.S..

The $e^{\pm}$ fluxes from $S$ decays are inevitably accompanied
by hard photons: coming from final state radiations (FSR) of cascade
decays, including $S\to\tilde{\tau}\to\tau\to\pi^{0}\to2\gamma$ and
the inverse Compton scattering (ICS) on the interstellar radiation
field (ISRF). The gamma ray fluxes could have Galactic and extragalactic
origins. We have calculated all these gamma ray spectra and compared
them with the recent Fermi LAT measurement in the region $0^{\circ}\leq l\leq360^{\circ},10^{\circ}\leq|b|\leq20^{\circ}$
\cite{Porter:2009sg}.

This paper is organized as follows. The supersymmetric SU(5) model
plus a singlet $S$ is presented in Section II, where we have also discussed
the possible decay channels of $S$ in some detail. In Section III,
a reasonable fit is obtained to reproduce the observed $e^{\pm}$
fluxes, by tuning relevant parameters in the model. Section IV
is devoted to the study of gamma-ray spectra from $e^{\pm}$ excesses.
Finally we conclude with a summary in section V. The component field
structure of the dimension six effective operators will be presented
in the Appendix. In this paper, we have used the Navarro-Frenk-White (NFW)
halo model \cite{Navarro:1996gj} for DM distribution and the MED propagation
model \cite{Donato:2003xg,Delahaye:2007fr}. For other halo and propagation
models, the conclusions are similar. In addition, all computations
on astrophysical effects are performed semi-analytically instead
of using the GALPROP program%
\footnote{Web page: http://galprop.stanford.edu/web\_galprop/galprop\_home.html%
}.

\section{a supersymmetric su(5) model}

If the observed $e^{\pm}$ excesses come from DM decays, the lifetime
of a TeV scale DM should be $\sim10^{26}s$. Such a long lifetime
can be naturally realized through decays via GUT suppressed dimension
six effective operators, similar to proton decays. This provides a
strong motivation to study DM decays in the framework of grand unification
theory \cite{Arvanitaki:2008hq,Arvanitaki:2009yb,Ruderman:2009ta,Ruderman:2009tj,Kadastik:2009cu,Kyae:2009gm}.

In the minimal supersymmetric SU(5) model, the dark matter candidate
would be the LSP, which is absolutely stable if R-parity is conserved.
In addition, the mass of LSP is normally around several hundred GeV,
which is too small to account for the Fermi and H.E.S.S. data even
if it decays. To make a minimal extension, one can introduce an SU(5) singlet $S$ as the dark matter candidate%
\footnote{If R parity is conserved, the neutralino LSP would also be part of
the DM. But for simplicity, we assume here that $S$ is the dominant
component of DM and the LSP contributes just a small portion to the
relic density. We will show that such a scenario is feasible in the next section.%
} \cite{Arvanitaki:2008hq}. As $S$ is neutral in the standard model
(SM) gauge group, it does not disturb the gauge coupling unification.
To eliminate lower dimensional operators which may lead $S$ to decay
too fast, we impose a $Z_{2}$ symmetry on the theory, under which
$S$ is odd while all other particles are even. Then $S$ can decay
into the MSSM particles only through dimension six operators, suppressed
by $M_{GUT}^{2}$. Assuming R-parity conservation and the $Z_{2}$
symmetry, all possible dimension six operators are \cite{Arvanitaki:2008hq}:
\begin{equation}
\frac{S^{+}S\overline{5}^{+}\overline{5}}{M_{GUT}^{2}},~~\frac{S^{+}STr(10^{+}10)}{M_{GUT}^{2}},~~\frac{S^{+}SW_{\alpha}W^{\alpha}}{M_{GUT}^{2}}~~\mbox{and}~~\frac{S^{+}SH_{u(d)}^{+}H_{u(d)}}{M_{GUT}^{2}}\label{eq:dimension-6 operators}\end{equation}
 Some or all of these operators may appear at the TeV scale when one
integrates out heavy particles of the GUT scale. Here $W_{\alpha}$
are the supersymmetric field strengths of SM gauge groups, $H_{u}$
and $H_{d}$ are the chiral superfields for Higgs, $\overline{5}$
and $10$ are anti-fundamental and antisymmetric tensor representations
of SU(5), respectively \begin{equation}
\overline{5}^{T}=(d^{c},d^{c},d^{c},e,-\nu)_{L}\end{equation}
 \begin{equation}
10=\frac{1}{\sqrt{2}}\left(\begin{array}{ccccc}
0 & u^{c} & -u^{c} & u & d\\
-u^{c} & 0 & u^{c} & u & d\\
u^{c} & -u^{c} & 0 & u & d\\
-u & -u & -u & 0 & e^{c}\\
-d & -d & -d & -e^{c} & 0\end{array}\right)_{L}\end{equation}

Operators $S^{+}SH_{u(d)}^{+}H_{u(d)}$ and $S^{+}SW_{\alpha}W^{\alpha}$
in Eq.(\ref{eq:dimension-6 operators}) may lead to final states containing
significant number of quarks or mono-energetic gamma ray lines. To
be consistent with experimental observations, these operators should
be further suppressed, most likely due to unknown physics at the GUT
scale. They will simply be neglected from now on. Operators
$S^{+}S\overline{5}^{+}\overline{5}$ and $S^{+}STr(10^{+}10)$ in
Eq.(\ref{eq:dimension-6 operators}) may be rewritten in the form
\begin{equation}
\underset{\Phi}{\sum}\frac{S^{+}S\Phi^{+}\Phi}{M_{GUT}^{2}}\label{eq:d6 operators in superfields}\end{equation}
 Here the summation is over all lepton and quark chiral superfields.
Assuming the singlet scalar develops a vacuum expectation value (VEV)
$<\widetilde{s}>$, the $Z_{2}$ symmetry is spontaneously broken
and both components $(\widetilde{s},s)$ in $S$ will decay.

Expanding Eq. (\ref{eq:d6 operators in superfields}) in terms of
component fields, one has \begin{equation}
\underset{\Phi}{\sum}\frac{1}{M_{GUT}^{2}}\left(i<\widetilde{s}>\widetilde{s}^{*}(\partial_{\mu}\psi\sigma^{\mu}\bar{\psi})+i<\widetilde{s}>\widetilde{\psi}^{*}(\partial_{\mu}\psi\sigma^{\mu}\bar{s})+<\widetilde{s}>\widetilde{s}^{*}\widetilde{\psi}^{*}\square\widetilde{\psi}\right)+h.c.+...~.\label{eq:3 operators}\end{equation}
 Here we have dropped total divergence terms. Operators from F-terms
have also been neglected as they are suppressed by the leptonic Yukawa
coupling constant. In addition, these operators will lead to many
body decays which are further suppressed by phase spaces. Details
of the expansion will be provided in Appendix \ref{appendix}.

To fit the $e^{\pm}$ fluxes data which steepens sharply above TeV,
the mass of $s$ and $\widetilde{s}$ will be assumed to be around
several TeV. In addition, the squark masses are assumed to be heavier
than the DM mass while the slepton masses to be about several hundred
GeV. So, $s$ and $\widetilde{s}$ can only decay into leptons, quarks
and sleptons, and have no squarks in the final state. The assumptions
on squark and slepton masses seems to be plausible, because squarks are much
heavier than sleptons in general. This is due to the fact that squarks
are color charged, which may affect drastically the renormalization
group equations for the squark masses.

The decay width of $s~$($\widetilde{s}$) due to the first two operators
in Eq.(\ref{eq:3 operators}) is proportional to the final state quark
or lepton mass square. For the first operator $\widetilde{s}^{*}(\partial_{\mu}\psi\sigma^{\mu}\bar{\psi})$,
the dominant decay channel is $\widetilde{s}\to t\bar{t}$, which
is suppressed by $M_{t}^{2}/M_{s}^{2}$ compared to the third operator
$\widetilde{s}^{*}\widetilde{\psi}^{*}\square\widetilde{\psi}$ in
Eq.(\ref{eq:3 operators}). The main decay channel of $s$ through
the second operator $\widetilde{\psi}^{*}(\partial_{\mu}\psi\sigma^{\mu}\bar{s})$
is $s\to\tau\widetilde{\tau}$, which is again suppressed by $M_{\tau}^{2}/M_{s}^{2}$
compared to the third operator in Eq.(\ref{eq:3 operators}). Thus
the decays of $s$ will not be considered. The DM $\widetilde{s}$
decays dominantly into a pair of sleptons, with universal coupling
for all generations. For simplification, we will simply neglect the
first two operators in Eq.(\ref{eq:3 operators}) and only consider
the operator $\widetilde{s}^{*}\widetilde{\psi}^{*}\square\widetilde{\psi}$
in the following. The remaining operator can be further rewritten
as \begin{equation}
\underset{\widetilde{l}}{\sum}\frac{-1}{M_{GUT}^{2}}<\widetilde{s}>\widetilde{s}^{*}(\widetilde{l_{L}}^{*}\square\widetilde{l_{L}}+\widetilde{l_{R}}\square\widetilde{l_{R}}^{*})\end{equation}
 with $\widetilde{l}=\widetilde{e},~\widetilde{\mu}$ and $\widetilde{\tau}$.
The corresponding decay width reads \begin{equation}
\Gamma_{\widetilde{l}}=\frac{\sqrt{M_{s}^{2}-4M_{\widetilde{l}}^{2}}<\widetilde{s}>^{2}M_{\widetilde{l}}^{4}}{16\pi M_{s}^{2}M_{GUT}^{4}}\end{equation}
 Taking $M_{GUT}=10^{16}$ GeV, $M_{s}\sim<\widetilde{s}>\sim$ a few
TeV and $M_{\widetilde{l}}\sim$ several hundred GeV, the lifetime
of $\widetilde{s}$ would be around $10^{26}s$, as one has hoped.
Notice also that the decay width is proportional to $M_{\widetilde{l}}^{4}$,
so that even slightly different masses between $\widetilde{e},~\widetilde{\mu}$
and $\widetilde{\tau}$ may lead to very different branching ratios.

With R-parity conservation, the slepton would decay to the LSP and
lepton quickly.%
\footnote{Here the LSP is assumed to be the neutralino. If the LSP is the gravitino
and $\widetilde{s}$ is heavier than $s$, $\widetilde{s}$ would
decay into $s$ predominately, instead of the sleptons.%
} $e^{\pm}$ can be produced through the following cascade decay chains:
\begin{itemize}
\item selectron chain: $\widetilde{s}\rightarrow\widetilde{e}\rightarrow e$
\item smuon chain: $\widetilde{s}\rightarrow\widetilde{\mu}\rightarrow\mu\rightarrow e$
\item stau chain: $\widetilde{s}\rightarrow\widetilde{\tau}\rightarrow\tau\rightarrow e$
\end{itemize}
In total, the $e^{\pm}$ fluxes due to DM decays at the source are
\begin{equation}
Q_{e}^{DM}(\vec{r},E)=\underset{\widetilde{l}}{\sum}\frac{\Gamma_{\widetilde{l}}^{DM}\rho^{DM}(\vec{r})}{M^{DM}}\frac{dN_{\widetilde{l}\rightarrow e}^{DM}}{dE}\label{eq:source}\end{equation}
 Here the summation is over all three cascade decay chains. $\Gamma_{\widetilde{l}}^{DM}$
is the decay width of the $\widetilde{l}$ cascade decay chain and
$M^{DM}$ is the DM mass. Since the lifetimes of sleptons, muon and
tau are extremely short compared with the DM decay, we can take the
approximation $\Gamma_{\widetilde{l}}^{DM}=\Gamma_{\widetilde{l}}$.
And $dN_{\widetilde{l}\rightarrow e}^{DM}/dE$ is the spectrum of
electron or positron per DM decay via a particular $\widetilde{l}$
chain. For the stau chain, the $e^{\pm}$ spectra are obtained by
using PYTHIA package \cite{Sjostrand:2006za}. $\rho^{DM}(r)$ is
the DM mass density which is model-dependent. As an illustration we
adopt the NFW halo model\cite{Navarro:1996gj} \begin{equation}
\rho^{DM}(r)=\frac{\rho_{\odot}r_{\odot}}{r}\left(\frac{1+r_{\odot}/r_{s}}{1+r/r_{s}}\right)^{2}\end{equation}
 with solar system position $r_{\odot}=8.5~\mbox{kpc}$, the DM density
at earth $\rho_{\odot}=0.3~\mbox{GeV}/\mbox{cm}^{3}$ and $r_{s}=20~\mbox{kpc}$.

\section{electron and positron excesses from dark matter decay}

\subsection{Positron and Electron Propagation}

Shown in Eq.(\ref{eq:source}) are the $e^{\pm}$ fluxes due to DM
decays at the source. However, only the $e^{\pm}$ fluxes at the Earth
are observable. It is thus necessary to consider the propagation of
electrons and positrons in the Galaxy. The $e^{\pm}$ flux per unit
energy at an arbitrary space-time point is given by \begin{equation}
\Phi_{e}^{DM}(t,\vec{r},E)=\frac{v_{e}}{4\pi}f_{e}^{DM}(t,\vec{r},E) .\end{equation}
For energetic $e^{\pm}$'s that we consider, their velocity $v_{e}$
is approximately equal to the light speed $c$. The function $f_{e}^{DM}(t,\vec{r},E)$
satisfies the diffusion-loss equation \begin{equation}
\frac{\partial f_{e}^{DM}}{\partial t}=K(E)\cdot\nabla^{2}f_{e}^{DM}+\frac{\partial}{\partial E}\left(B(E)f_{e}^{DM}\right)+Q_{e}^{DM}~.\label{eq:diffusion}\end{equation}
 Here the convection and advection terms have been neglected. $Q_{e}^{DM}$
is due to the DM decays as given in Eq.(\ref{eq:source}). $K(E)$
stands for the diffusion coefficient which is related to the rigidity
of the particle. For $e^{\pm}$, it can be parameterized as \begin{equation}
K(E)=K_{0}(E/GeV)^{\delta}~.\end{equation}
 $B(E)=E^{2}/(\mbox{GeV}\cdot\tau_{E})$ is the effective energy loss
coefficient with $\tau_{E}=10^{16}s$, which describes the energy
loss of $e^{\pm}$ due to ICS on the ISRF and synchrotron radiation.
Eq. (\ref{eq:diffusion}) can be solved in a solid flat cylinder parameterized
by $(r,z,\theta)$, with $z\in[-L,L]$ in z direction and $r\in[0,~20~\mbox{kpc}]$
in radius. The solar system corresponds to the position $(r_{\odot},z_{\odot},\theta_{\odot})=(8.5~\mbox{kpc},~0,~0)$.
$f_{e}^{DM}(t,\vec{x},E)$ is assumed to vanish on the surface of
the flat cylinder, which serves as the boundary condition for this
equation. For the MED propagation model, $L$, $\delta$ and $K_{0}$
are chosen to be 4~kpc, 0.70 and $0.0112~\mbox{kpc}^{2}/\mbox{Myr}$,
respectively \cite{Cirelli:2008id}.

If $f_{e}^{DM}(t,\vec{x},E)$ does not change with time, a steady
state solution Eq. (\ref{eq:diffusion}) can be obtained semi-analytically
\cite{Delahaye:2007fr,Cirelli:2008id,Cheung:2009si,Ishiwata:2009vx}.
The $e^{\pm}$ fluxes at the Earth are \begin{equation}
\Phi_{e}^{DM}(r_{\odot},E)=\frac{c}{4\pi B(E)}\underset{\widetilde{l}}{\sum}\frac{\rho_{\odot}\Gamma_{\widetilde{l}}^{DM}}{M^{DM}}\int_{E}^{M^{DM}/2}dE'I(\lambda_{D}(E,E'))\frac{dN_{\widetilde{l}\rightarrow e}^{DM}}{dE'}~.\label{eq:dm flux}\end{equation}
 Here $\lambda_{D}(E,E')$ describes the diffusion length from energy
$E'$ to $E$, which can be parameterized as \begin{equation}
\lambda_{D}^{2}=4K_{0}\tau_{E}\left(\frac{(E/\mbox{GeV})^{\delta-1}-(E'/\mbox{GeV})^{\delta-1}}{1-\delta}\right)~.\end{equation}
 The function $I(\lambda_{D})$ is given by: \begin{equation}
I(\lambda_{D})=a_{0}+a_{1}\tanh\left(\frac{b_{1}-l}{c_{1}}\right)\left[a_{2}\exp\left(-\frac{(l-b_{2})^{2}}{c_{2}}\right)+a_{3}\right]\label{eq:I}\end{equation}
 with $l=\log_{10}(\lambda_{D}/\mbox{kpc})$. $I(\lambda_{D})$ contains
the whole information of the NFW halo model and MED propagation model.
Parameters in Eq.(\ref{eq:I}) have been estimated numerically
in Ref. \cite{Cirelli:2008id} and listed in Table \ref{tab:1}.

\begin{table}
\begin{tabular}{|c|c|c|c|c|c|c|c|}
\hline
$a_{0}$  & $a_{1}$  & $a_{2}$  & $a_{3}$  & $b_{1}$  & $b_{2}$  & $c_{1}$  & $c_{2}$\tabularnewline
\hline
\hline
0.502  & 0.621  & 0.688  & 0.806  & 0.891  & 0.721  & 0.143  & 0.071\tabularnewline
\hline
\end{tabular}\caption{\label{tab:1}Parameters in Eq.(\ref{eq:I}) in accord with the NFW
halo model and MED propagation model.}

\end{table}

\subsection{Positron and Electron Backgrounds}

For interstellar background fluxes of $e^{\pm}$, we use the {}``model
0'' proposed by the Fermi LAT collaboration\cite{Grasso:2009ma},
which can be parameterized as \cite{Ibarra:2009dr}. \begin{eqnarray}
\Phi_{e^{-}}^{bkg}(E) & = & \frac{82.0\epsilon^{-0.28}}{1+0.224\epsilon^{2.93}}~,\\
\Phi_{e^{+}}^{bkg}(E) & = & \frac{38.4\epsilon^{-4.78}}{1+0.0002\epsilon^{5.63}}+24.0\epsilon^{-3.41}\end{eqnarray}
 in units of $\mbox{GeV}^{-1}\mbox{m}^{-2}\mbox{s}^{-1}\mbox{sr}^{-1}$,
where $\epsilon=E/1\mbox{GeV}$.

For electron/positron fluxes at the top of the Earth's atmosphere
$\Phi_{e^{\pm}}^{\oplus}$, solar modulation effects should be considered.
Adopting the force field approximation, one has \cite{Perko:1987pq}
\begin{equation}
\Phi_{e^{\pm}}^{\oplus}(E_{\oplus})=\frac{E_{\oplus}^{2}}{E_{IS}^{2}}\Phi_{e^{\pm}}^{IS}(E_{IS})~,\end{equation}
 where $\Phi_{e^{\pm}}^{IS}$ stand for interstellar fluxes and $E_{\oplus}=E_{IS}+|Ze|\phi_{F}$, with $\phi_{F}=0.55$GV as
a typical value. It is clear that, at energies larger than $10$GeV,
solar modulation effects could be neglected as $E_{\oplus}\thickapprox E_{IS}$.

Finally, the $e^{\pm}$ fluxes and positron fraction at the top of
the Earth's atmosphere could be expressed as \begin{equation}
E_{\oplus}^{3}\times\Phi_{\oplus}(E_{\oplus})=E_{\oplus}^{3}\times\frac{E_{\oplus}^{2}}{E_{IS}^{2}}[\Phi_{e^{+}}^{bkg}(E_{IS})+\Phi_{e^{+}}^{DM}(E_{IS})+N\cdot\Phi_{e^{-}}^{bkg}(E_{IS})+\Phi_{e^{-}}^{DM}(E_{IS})]~,\label{eq:total flux}\end{equation}
 \begin{equation}
\frac{\Phi_{e^{+}}(E_{\oplus})}{\Phi_{e^{+}}(E_{\oplus})+\Phi_{e^{-}}(E_{\oplus})}=\frac{\Phi_{e^{+}}^{bkg}(E_{IS})+\Phi_{e^{+}}^{DM}(E_{IS})}{\Phi_{e^{+}}^{bkg}(E_{IS})+\Phi_{e^{+}}^{DM}(E_{IS})+N\cdot\Phi_{e^{-}}^{bkg}(E_{IS})+\Phi_{e^{-}}^{DM}(E_{IS})}~.\label{eq:fraction}\end{equation}
 Here $N$ is a normalization factor standing for the uncertainty
of the electron flux. In this paper $N=0.8$ is chosen to fit the
experimental data.

\subsection{A Fit of PAMELA and Fermi LAT Data}

\begin{figure}[tb]
 \includegraphics[bb=50bp 25bp 510bp 360bp,clip,width=3.2in]{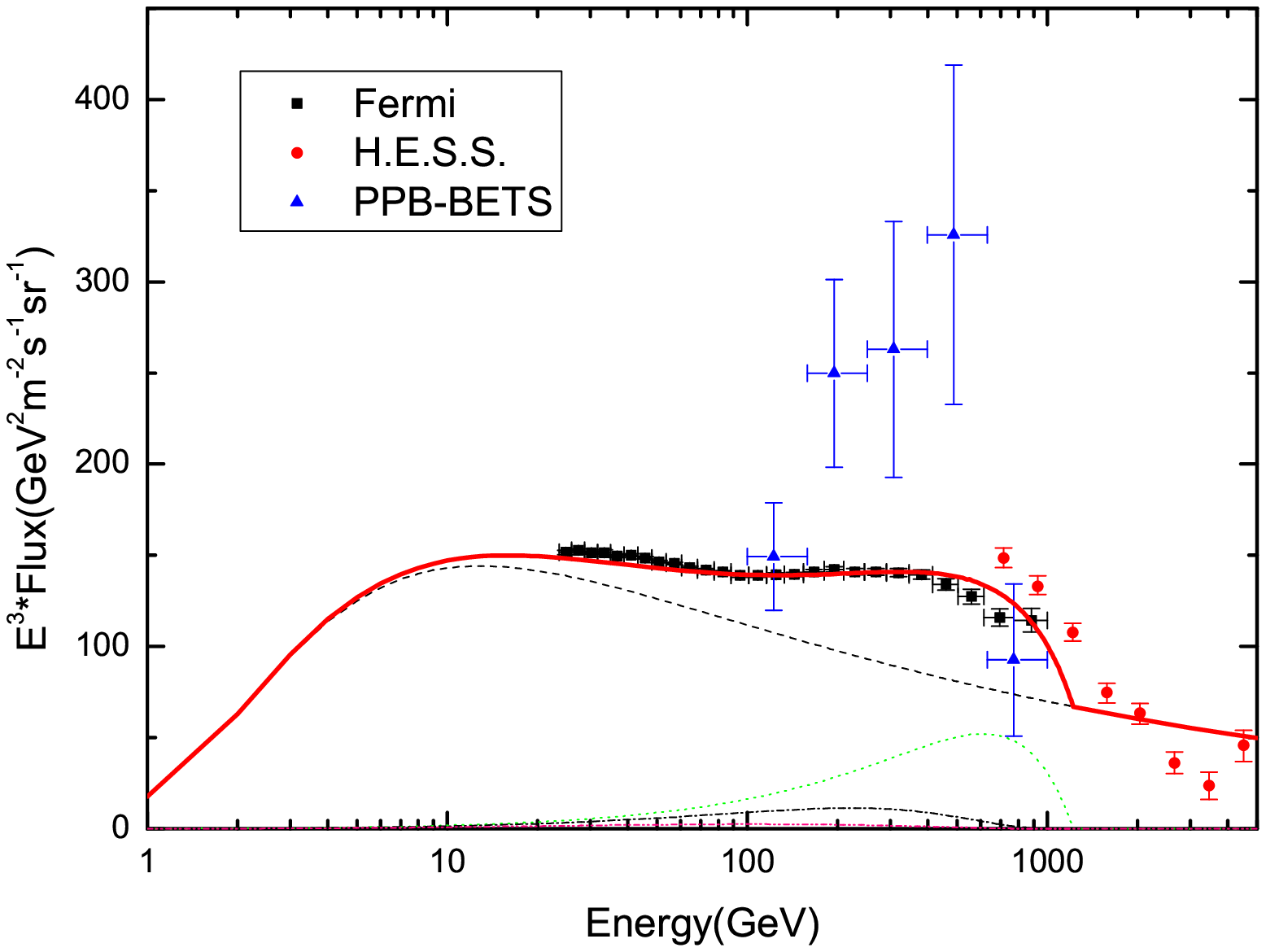}
\includegraphics[bb=50bp 25bp 510bp 360bp,clip,width=3.2in]{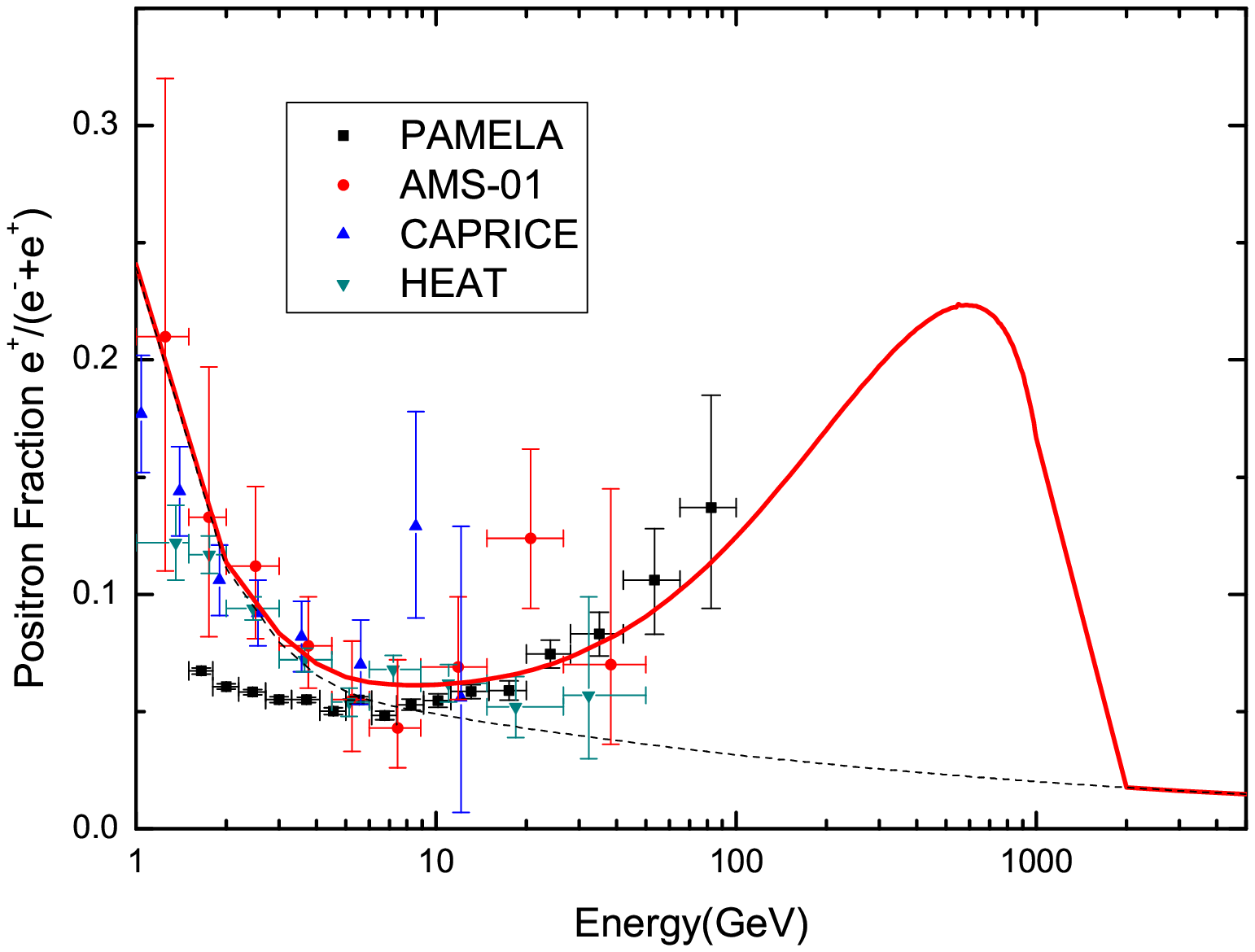}
\caption{The SU(5) model gives a reasonable fit to the Fermi, H.E.S.S. and Pamela data with the example set of
parameters given in the text.
Left: Decaying DM produces extra $e^- + e^+$
fluxes above background via three different cascade decay chains.
The green dot line, black dash dot line and pink dash dot dot line represent the selectron chain, the smuon chain
and the stau chain, respectively. The red solid line
includes all contributions from our fit. The black dash line shows the background as discussed
in the text.
Right: Including the $e^{+}$ background flux, decaying
DM predicts a positron fraction which fits the experimental data.
The red solid line shows the result of our fit, and the black dash line indicates
the background.}
\label{fig:fit}
\end{figure}

For illustration, we choose DM mass $M_{\widetilde{s}}=6.5$ TeV,
$M_{GUT}=10^{16}$ GeV, $<\widetilde{s}>=20$ TeV, $M_{\widetilde{e}}=380$
GeV, $M_{\tilde{\mu}}=370$ GeV, $M_{\widetilde{\tau}}=330$ GeV and
$M_{LSP}=300$ GeV.%
\footnote{We have checked explicitly that, for this set of slepton masses, the neutrolino LSP could be only a minor part of the DM. For
example, by using DarkSUSY package\cite{Gondolo:2004sc}, we have obtained
$\Omega_{LSP}h^{2}=0.009$, with the neutralino mass spectrum $M_{LSP}=M_{\chi_{1}}=300$
GeV, $M_{\chi_{2}}=315$ GeV, $M_{\chi_{3}}=630$ GeV, $M_{\chi_{4}}=690$
GeV and the gaugino fraction to be $0.03$. %
} With this parameter set, the decaying DM produces extra $e^{\pm}$
fluxes from 10 GeV to 1 TeV, as shown in Fig.\ref{fig:fit}a. The
main contribution comes from the selectron chain. The cascade decay
$\widetilde{s}\to\widetilde{e}^{+}+\widetilde{e}^{-}\to e^{+}+e^{-}$
smoothes the $e^{+}+e^{-}$ spectrum and naturally allows for a good
fit to the Fermi LAT measurement. The $e^{\pm}$ fluxes steepen above
1TeV sharply, which is consistent with the H.E.S.S. observation. The
positron fraction are shown in Fig.\ref{fig:fit}b, compared to
the data of PAMELA, AMS-01, CAPRICE and HEAT.

\section{diffuse gamma-rays from the $e^{\pm}$ excesses}

The $e^{\pm}$ excesses are inevitably accompanied by photons coming
from the FSR, ICS and synchrotron radiation stemming from them.

(1) FSR: The bremsstrahlung of $e^{\pm}$ fluxes leads to the emission
of energetic photon flux $\Phi_{FSR}$. Moreover, in our model the
stau chain contains $\tau$ lepton which emits hard photons via the
process $\tau\rightarrow\pi^{0}\rightarrow\gamma+\gamma$. This mechanism
is significant, especially at the high energy end of the spectrum.
The largest energy of FSR photons could be around $M^{DM}/2$, which
could be probed by the H.E.S.S. collaboration. Notice also that the
spectrum of FSR is quite model-dependent. In addition, the FSR could
come from within or without our Galaxy ($\Phi_{FSR}^{GAL}/\Phi_{FSR}^{EG}$).

(2) ICS: The ICS radiation is produced when the $e^{\pm}$ excesses
scatter on the ISRF. In our Galaxy the ISRF includes the cosmic microwave
background (CMB), star light and the infrared light, while outside
of our Galaxy the CMB component is dominant. The corresponding
ICS fluxes $\Phi_{ICS}^{GAL}$ and $\Phi_{ICS}^{EG}$, which should
be observable by Fermi LAT, are closely related to the $e^{\pm}$
excesses. Therefore the ICS spectrum is, to some extent, model independent
as long as the DM model can reproduce the
Fermi and H.E.S.S. $e^{\pm}$ spectrum with reasonable accuracies.

(3) Synchrotron radiation: During propagation, the $e^{\pm}$ fluxes
radiate photons in the Galactic magnetic fields. These photons should
be very soft (around $10^{-6}\mbox{eV}$) and outside the energy range
explored by Fermi LAT and H.E.S.S. experiments. They will not be considered
in the following.

Notice that the extragalactic gamma rays are roughly of the same order
as the Galactic ones. But the extragalactic component is isotropic
while the Galactic one has angular dependence. The total gamma ray
flux is obtained by summing all these contributions: \begin{equation}
\Phi_{\gamma}=\Phi_{FSR}^{GAL}+\Phi_{ICS}^{GAL}+\Phi_{FSR}^{EG}+\Phi_{ICS}^{EG}\end{equation}
 Specifically, we consider only photons in the region $0^{\circ}\leq l\leq360^{\circ},~10^{\circ}\leq|b|\leq20^{\circ}$
in the following, as Fermi LAT has released the data in this region
recently \cite{Porter:2009sg}.

\subsection{Galactic Gamma Rays from FSR}

As photons propagate almost freely in the Galaxy, the differential
flux of photons received at the Earth in a given solid angle $d\Omega$
is given by \cite{Cirelli:2008id}

\begin{equation}
\frac{d\Phi_{FSR}^{GAL}}{dE_{\gamma}d\Omega}=2\frac{r_{\odot}}{4\pi}\frac{\rho_{\odot}}{M^{DM}}\overline{J}\underset{\widetilde{l}}{\sum}\Gamma_{\widetilde{l}}
\frac{dN_{\widetilde{l}\rightarrow\gamma}^{DM}}{dE_{\gamma}}.\label{eq:galactic fsr}\end{equation}
Here the factor of 2 takes into account the fact that both leptons
and anti-leptons contribute equally to the FSR flux of gamma rays.
$dN_{\widetilde{l}\rightarrow\gamma}^{DM}/dE_{\gamma}$ is the photon
spectrum per DM decay via a specific slepton chain. PYTHIA package
\cite{Sjostrand:2006za} has been used here to obtain these spectra.
$\overline{J}$ encodes all the astrophysical information which is
defined as \begin{equation}
\overline{J}\triangle\Omega=\int d\Omega\int_{0}^{\infty}\frac{ds}{r_{\odot}}\frac{\rho(r)}{\rho_{\odot}}~,\label{eq:j}\end{equation}
 where the parameter $s$ is integrated along a line of sight. The
parameter $s$ can be linked to parameters $r$, $l$ and $b$ by
\begin{equation}
r(s,l,b)=\sqrt{s^{2}+r_{\odot}^{2}-2sr_{\odot}\cos l\cos b}\end{equation}

\begin{figure}[tb]
\includegraphics[bb=50bp 25bp 510bp 370bp,clip,width=3.2in]{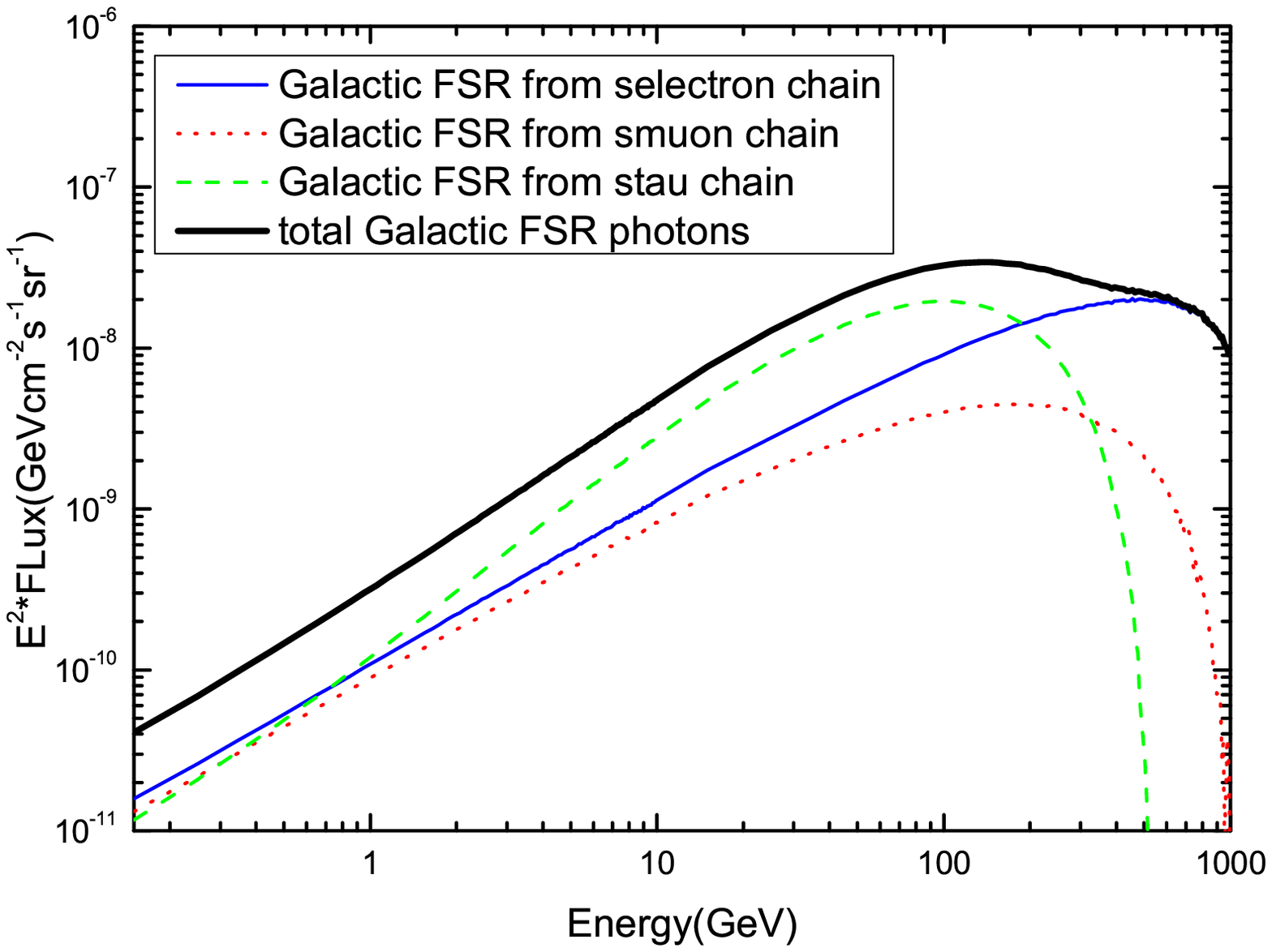}
\includegraphics[bb=50bp 25bp 510bp 370bp,clip,width=3.2in]{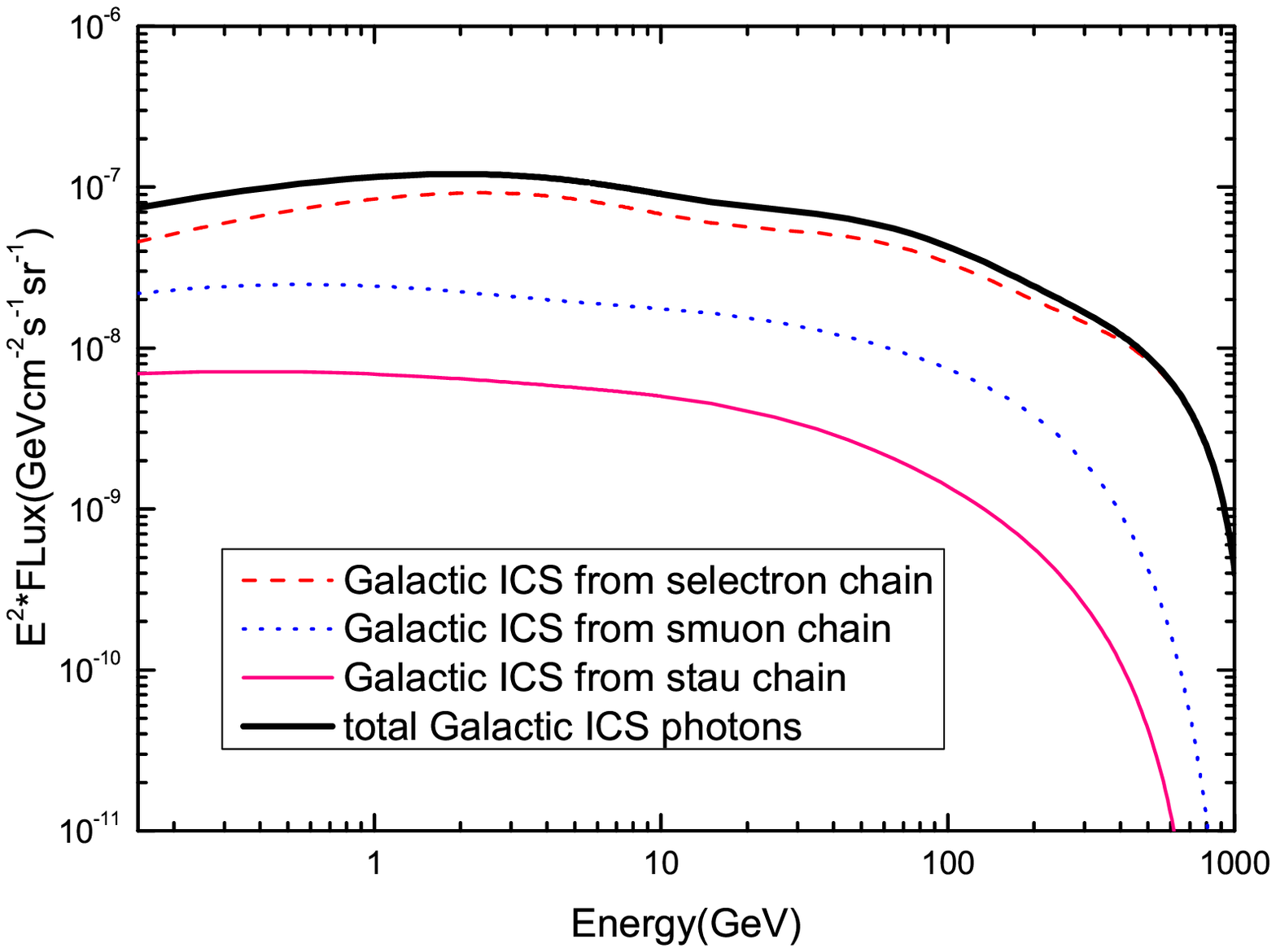}
\caption{Galactic gamma ray spectra from FSR (left) and ICS (right) via three different decay chains, plotted
in the region $0^{\circ}\leq l\leq360^{\circ},~10^{\circ}\leq|b|\leq20^{\circ}$.}
\label{fig:galactic}
\end{figure}

From Eq.(\ref{eq:j}), one obtains $\overline{J}=2.4$ in the region
$0^{\circ}\leq l\leq360^{\circ},~10^{\circ}\leq|b|\leq20^{\circ}$.
The photon spectra from Galactic FSR are plotted in Fig.\ref{fig:galactic}a, which
peak around several hundred GeV. Notice that the stau chain gives
a large contribution to the photon spectrum due to $\tau\rightarrow\pi^{0}\rightarrow\gamma+\gamma$.

\subsection{Galactic Gamma Rays from ICS }

A pedagogical review about ICS was provided in \cite{Blumenthal:1970gc}.
We will calculate the ICS gamma rays semi-analytically, following Refs.
\cite{Cirelli:2009vg,Ibarra:2009dr,Ibarra:2009nw,Ishiwata:2009dk}.

The differential flux of ICS photons received at the Earth in a given
solid angle $d\Omega$ with energy between $E_{\gamma}$ and $E_{\gamma}+dE_{\gamma}$
can be expressed as: \begin{equation}
\frac{d\Phi_{ICS}^{GAL}}{dE_{\gamma}d\Omega}=\frac{2c}{4\pi\triangle\Omega}\int d\Omega\int_{0}^{\infty}ds\int_{0}^{\infty}d\epsilon\int_{M_{e}}^{M^{DM}/2}dE_{e}\frac{d\sigma^{ICS}(E_{e},\epsilon)}{dE_{\gamma}}f_{e}^{DM}(\vec{r},E_{e})f_{ISRF}(\vec{r},\epsilon)~,\label{eq:ICS1}\end{equation}
here $f_{e}(\vec{r},E_{e})$ denotes initial electron number density
and $f_{ISRF}(\vec{r},\epsilon)$ the ISRF photon number density.
The factor of 2 reflects that both electrons and positrons contribute
to the ICS gamma rays equally. The Compton cross section is given
by the Klein-Nishina formula \begin{equation}
\frac{d\sigma^{ICS}(E_{e},\epsilon)}{dE_{\gamma}}=\frac{3\sigma_{T}}{4\gamma_{e}^{2}\epsilon}\left(2q\ln q+1+q-2q^{2}+\frac{(q\Gamma)^{2}}{2(1+q\Gamma)}(1-q)\right)\end{equation}
 where \begin{equation}
q=\frac{E_{\gamma}}{\Gamma(E_{e}-E_{\gamma})},~\Gamma=\frac{4\gamma_{e}\epsilon}{m_{e}},~\gamma_{e}=\frac{E_{e}}{m_{e}}~.\end{equation}
 Here $\sigma_{T}=0.67$ barn is the Compton scattering cross section
in the Thomson limit and $m_{e}$ is the electron mass. Kinematics
requires that $\epsilon\leq E_{\gamma}\leq(1/E_{e}+1/4\gamma_{e}^{2}\epsilon)^{-1}$.

The initial electron or positron number density $f_{e}(\vec{r},E_{e})$
can be obtained by solving Eq.(\ref{eq:diffusion}) at each position.
Notice that Eq.(\ref{eq:diffusion}) is dominated by the energy loss
term at high energy. That is to say, $e^{\pm}$ can not propagate
far from the production position before losing most of their energy.
Therefore, Eq.(\ref{eq:diffusion}) may be solved point by point approximately
\begin{equation}
f_{e}^{DM}(\vec{r},E_{e})=\frac{1}{B(E_{e})}\frac{\rho(\vec{r})}{M^{DM}}\underset{\widetilde{l}}{\sum}\Gamma_{\widetilde{l}}Y_{\widetilde{l}}(E_{e})\end{equation}
 with \begin{equation}
Y_{\widetilde{l}}(E_{e})=\int_{E_{e}}^{M^{DM}/2}dE'\frac{dN_{\widetilde{l}\rightarrow e}^{DM}}{dE'}~.\label{eq:Y function}\end{equation}

To approximate further, we will assume that the ISRF photons have
the same energy spectra at any point in the region $0^{\circ}\leq l\leq360^{\circ},~10^{\circ}\leq|b|\leq20^{\circ}$.
That is to say, the number density of ISRF $f_{ISRF}(\vec{r},\epsilon)=f_{ISRF}(\epsilon)$,
which can be described by three blackbody-like spectra roughly \cite{Cirelli:2009vg}:
\begin{equation}
f_{ISRF}(\epsilon)=\underset{i=1,2,3}{\sum}N_{i}\frac{\epsilon^{2}}{\pi^{2}}\frac{1}{e^{\epsilon/T_{i}}-1}\end{equation}
 with $T_{1}=2.753$ K, $N_{1}=1$ for the CMB, $T_{2}=3.5\times10^{-3}$
eV, $N_{2}=1.3\times10^{-5}$ for the infrared light and $T_{3}=0.3$
eV, $N_{3}=8.9\times10^{-13}$ for the star light.

In order to separate astrophysics and particle physics information,
Eq.(\ref{eq:ICS1}) can be rewritten as \begin{equation}
\frac{d\Phi_{ICS}^{GAL}}{dE_{\gamma}d\Omega}=2c\overline{J}\frac{r_{\odot}}{4\pi}\frac{\rho_{\odot}}{M^{DM}}\underset{\widetilde{l}}{\sum}\int_{0}^{\infty}d\epsilon\int_{M_{e}}^{M^{DM}/2}dE_{e}\frac{d\sigma^{ICS}(E_{e},\epsilon)}{dE_{\gamma}}\frac{1}{B(E_{e})}\Gamma_{\widetilde{l}}Y_{\widetilde{l}}(E_{e})f_{ISRF}(\epsilon)~.\end{equation}
Shown in Fig.\ref{fig:galactic}b is the ICS photon spectra in our model. One
sees that the gamma ray fluxes come mostly from the selectron chain
and steepen sharply above $1$ TeV.

\subsection{Extragalactic Gamma Rays from FSR}

\begin{figure}[tb]
\includegraphics[bb=50bp 25bp 510bp 370bp,clip,width=3.2in]{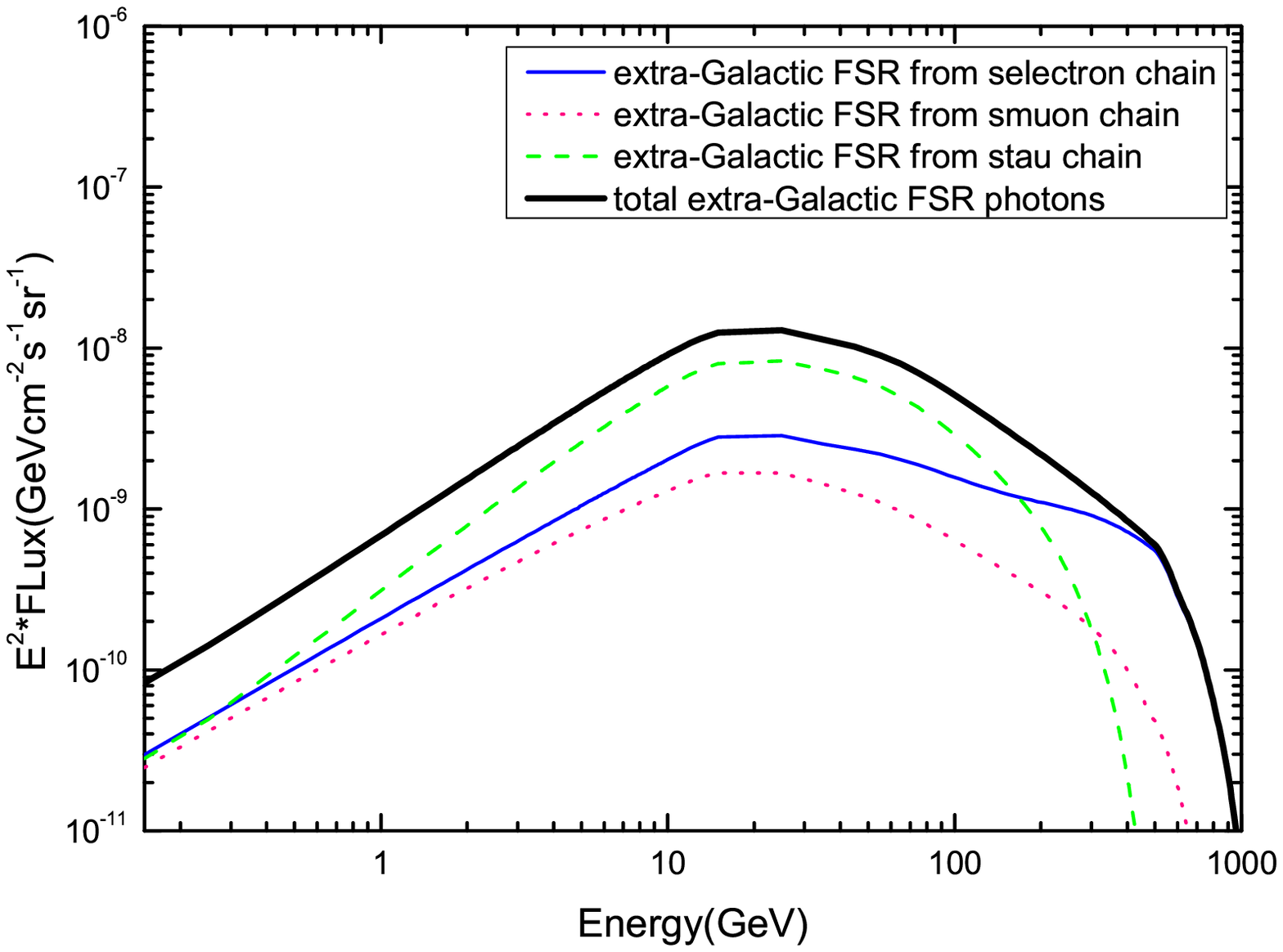}
\includegraphics[bb=50bp 25bp 510bp 370bp,clip,width=3.2in]{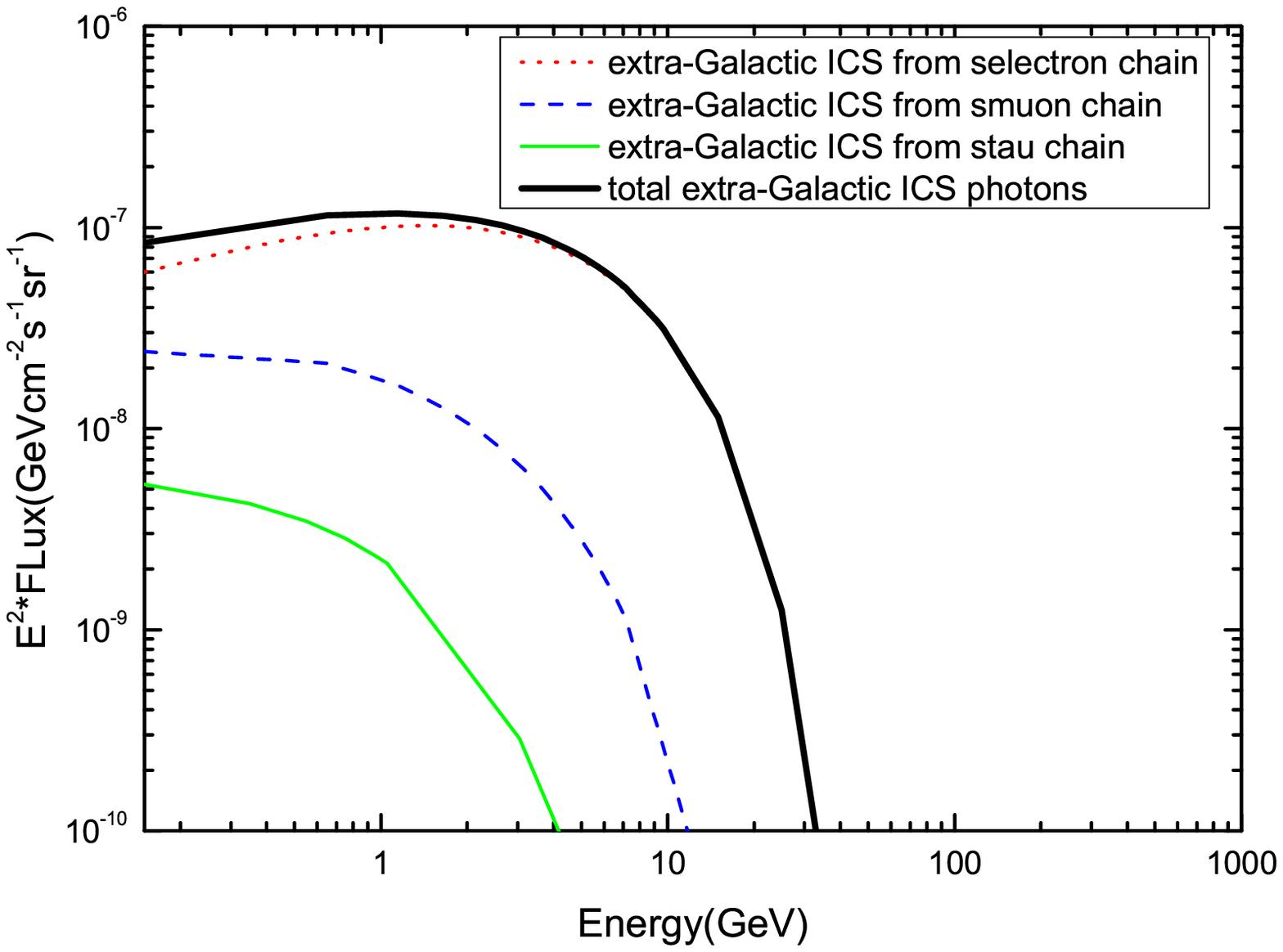}
\caption{Extragalactic gamma ray spectra from FSR (left) and ICS (right) via three different decay chains,
plotted in the region $0^{\circ}\leq l\leq360^{\circ},~10^{\circ}\leq|b|\leq20^{\circ}$.}
\label{fig:extra-galactic}
\end{figure}

To study gamma rays from the outside of our Galaxy, the effects
due to the expansion of the Universe should be considered. By turning
the line-of-sight integral into a redshift integral, the differential
flux of isotropic photons of the extragalactic origin is given by
\cite{Ibarra:2009nw,Chen:2009uq} \begin{equation}
\frac{d\Phi_{FSR}^{EG}}{dE_{\gamma}d\Omega}=\frac{2c}{4\pi}\frac{\Omega_{DM}\rho_{c}}{M_{DM}}\underset{\widetilde{l}}{\sum}\Gamma_{\widetilde{l}}\int_{0}^{\infty}dz\frac{e^{-\tau(E_{\gamma},z)}}{H(z)}\left.\frac{dN_{\widetilde{l}\rightarrow\gamma}^{DM}}{dE'_{\gamma}}\right|_{E'_{\gamma}=(1+z)E_{\gamma}}~.\end{equation}
 Here $z$ is the redshift, $H(z)=H_{0}\sqrt{\Omega_{\Lambda}+\Omega_{M}(z+1)^{3}}$
is the Hubble expansion rate, with $H_{0}=100\mbox{h km}\mbox{s}^{-1}\mbox{Mpc}^{-1}$
and the present day normalized Hubble expansion rate $h=0.72$\cite{Dunkley:2008ie}.
$\rho_{c}=5.5\times10^{-6}\mbox{GeV}/\mbox{cm}^{3}$ is the critical
density of the Universe. We also take the dark matter density $\Omega_{DM}=0.21$,
the dark energy density $\Omega_{\Lambda}=0.74$ and the matter density
$\Omega_{M}=0.26$ \cite{Dunkley:2008ie}. The spectrum $dN_{\widetilde{l}\rightarrow\gamma}^{DM}/dE_{\gamma}'$
is the same as that in Eq (\ref{eq:galactic fsr}), except that the
redshift effect has been included. The parametric form for the optical
depth $\tau(E_{\gamma},z)$ of the \textquotedbl{}fast evolution\textquotedbl{}
model could be found in Refs. \cite{Stecker:2005qs,Stecker:2006eh}.
Fig.\ref{fig:extra-galactic}a shows those contributions of our model.
Again the stau chain is important here because of the $\pi^{0}$ channel.

\subsection{Extragalactic Gamma Rays from ICS }

We adopt a semi-analytical calculation following Refs.\cite{Ishiwata:2009dk,Chen:2009uq}.
Concerning the dilution effect due to the expansion of the Universe,
the diffusion-loss equation of electrons and positrons becomes \begin{equation}
\frac{\partial f_{e}^{DM}}{\partial t}=H(z)E_{e}\frac{\partial f_{e}^{DM}}{\partial E_{e}}+\frac{\partial}{\partial E_{e}}\left[B^{EG}(z,E_{e})f_{e}^{DM}\right]+Q_{e}^{DM}~.\label{eq:extra-diffusion}\end{equation}
 Here the extragalactic energy loss rate $B^{EG}(z,E)$ is given as \begin{equation}
B^{EG}(z,E_{e})=\frac{4}{3}\sigma_{T}\gamma_{e}^{2}\rho_{CMB}(1+z)^{4}\end{equation}
 with $\rho_{CMB}=\Omega_{\gamma}\rho_{c}=0.26\times10^{-9}\mbox{GeV}/\mbox{cm}^{3}$
the present-day CMB energy density. For $e^{\pm}$ energy around several
hundred GeV, the timescale of energy-loss is $E/B^{EG}(z,E)\sim10^{14}s$,
which is much less than the Hubble time. That is to say, basically
$e^{\pm}$ do not feel the redshift effect before losing most of their
energy. So the Hubble term in Eq(\ref{eq:extra-diffusion}) can be
safely neglected. The $e^{\pm}$ spectrum from DM decay can then be
solved as\begin{equation}
f_{e}^{DM}(z,E_{e})=\frac{(1+z)^{3}}{B^{EG}(z,E_{e})}\frac{\Omega_{DM}\rho_{c}}{M^{DM}}\underset{\widetilde{l}}{\sum}\Gamma_{\widetilde{l}}^{DM}Y_{\widetilde{l}}(E_{e})~.\end{equation}
 Finally, the differential flux of extragalactic ICS photons received
at the Earth in an arbitrary solid angle $d\Omega$ with energy between
$E_{\gamma}$ and $E_{\gamma}+dE_{\gamma}$ can be expressed as: \begin{equation}
\frac{d\Phi_{ICS}^{EG}}{dE_{\gamma}d\Omega}=\frac{2c}{4\pi}\int_{0}^{\infty}\frac{dzd\epsilon dE_{e}}{(1+z)^{3}H(z)}f_{\gamma}(z,\epsilon)f_{e}^{DM}(z,E_{e})\left.\frac{d\sigma^{ICS}(z,E_{e},\epsilon)}{dE'_{\gamma}}\right|_{E'_{\gamma}=(1+z)E_{\gamma}}~.\end{equation}
 The spectrum $f_{\gamma}(z,\epsilon)$ of the background CMB radiation
at redshift $z$ is given as\begin{equation}
f_{\gamma}(z,\epsilon)=\frac{\epsilon^{2}}{\pi^{2}}\frac{1}{e^{\epsilon/[(1+z)T]}-1}\end{equation}
 with $T=2.753$ K. The photon spectra from extragalactic ICS are
plotted in Fig.\ref{fig:extra-galactic}b, which are dominated by the selectron
chain contribution. Because of the redshift, the spectrum drops rapidly
at high energy.

\begin{figure}
\includegraphics[scale=0.7]{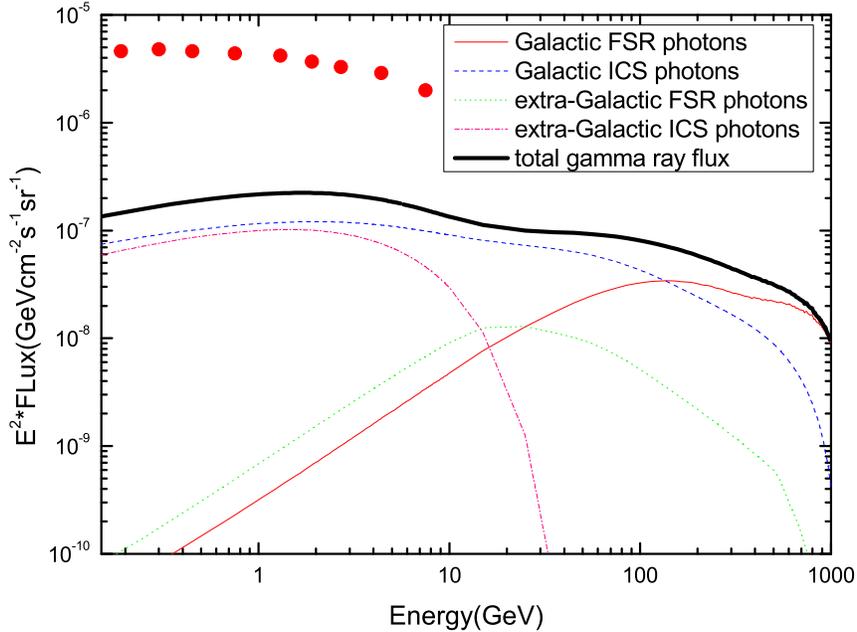} \caption{\label{fig:total}All gamma ray spectra in the region $0^{\circ}\leq l\leq360^{\circ},10^{\circ}\leq|b|\leq20^{\circ}$.
Dots stand for the preliminary Fermi LAT data \cite{Porter:2009sg}. }

\end{figure}

Finally, Fig. \ref{fig:total} shows the total gamma ray spectra including
all contributions. One can see that the FSR $\gamma$-rays dominate
at higher energies while the ICS ones dominate at lower energies.
Extragalactic gamma rays are not significant at high energy due to
the redshift effect. The total gamma ray spectrum from $e^{\pm}$
excesses are consistent with the preliminary Fermi LAT data \cite{Porter:2009sg}
from $0.1$ GeV to $10$ GeV, as shown in Fig. \ref{fig:total}. The
predicted gamma ray flux around several hundred GeV may be tested
by the Fermi satellite in the near future.

\section{summary}

In this paper we have studied the DM decay in supersymmetric SU(5)
models. An SU(5) singlet $S$, instead of LSP, is assumed to be the
dominant component of DM. With R-parity
conservation and a spontaneously broken $Z_{2}$ symmetry, the singlet
$S$ can decay into visible particles through dimension six effective
operators suppressed by the GUT scale. Assuming the squarks
to be heavier than $S$, $S$ decays dominantly into a pair
of sleptons through the effective operator $\widetilde{s}^{*}(\widetilde{l_{L}}^{*}\square\widetilde{l_{L}}+\widetilde{l_{R}}\square\widetilde{l_{R}}^{*})$.
Typically, the lifetime of $S$ is around $10^{26}$ s, much longer
than the age of the Universe. Since the decay products of $S$ do
not contain any quarks, our model is consistent with the Pamela antiproton
measurement automatically.
For illustration, we have chosen $M^{DM}=6.5$
TeV, $M_{GUT}=10^{16}$ GeV, $M_{\widetilde{e}}=380$ GeV, $M_{\tilde{\mu}}=370$
GeV, $M_{\widetilde{\tau}}=330$ GeV and $M_{LSP}=300$ GeV. With
this parameter set, we have shown that the $S$ decays can account
for the PAMELA, H.E.S.S. and Fermi LAT $e^{\pm}$ excesses. Numerically,
we have adopted the NFW profile for the dark matter distribution and
the MED propagation model for the cosmic ray propagation. To interpret
these results, one should keep in mind that there exist substantial
astrophysical uncertainties about $e^{\pm}$ background, $e^{\pm}$
propagation and the DM distribution.

When $e^{\pm}$ propagate from the decay position to the
Earth, hard photons are emitted inevitable due to inverse Compton
scattering and final state radiation. Future measurements of the diffuse
gamma ray may distinguish DM explanations from astrophysical explanations
by looking into the energy and angular distributions. We have calculated
the gamma ray spectra in our model. The predicted photon spectrum
are compared to the preliminary Fermi LAT measurement from $0.1$
GeV to $10$ GeV in the region $0^{\circ}\leq l\leq360^{\circ},~10^{\circ}\leq|b|\leq20^{\circ}$,
which seems to be consistent with each other. The total gamma ray
spectrum are dominated by photons from Galactic final state radiation
for the photon energy above $100$ GeV, which may be tested by Fermi
LAT in the near future.
\begin{acknowledgments}
We thank Xiao-Jun Bi for helps on astrophysical matters and Yufeng Zhou for
useful discussions on diffuse gamma rays. This work is supported in
part by the National Science Foundation of China (No.10875103, No.10425525 and No. 10705024)
and National Basic Research Program of China (2010CB833000). G.Z is
also supported in part by Chinese Universities Scientific Fund and
in part by the Scientific Research Foundation for the Returned Overseas
Chinese Scholars, State Education Ministry.
\end{acknowledgments}
\appendix

\section{the component field structure of $S^{+}S\Phi^{+}\Phi$}

\label{appendix} We now provide the component field structure of the dimension
six operator $S^{+}S\Phi^{+}\Phi$. Define \begin{equation}
\Phi(y)=\tilde{\psi}(y)+\sqrt{2}\theta\psi(y)+\theta^{2}F_{\psi}(y),~~\Phi^{+}(y^{+})=\tilde{\psi}^{*}(y^{+})+\sqrt{2}\bar{\theta}\bar{\psi}(y^{+})+\bar{\theta}^{2}F_{\psi}^{*}(y^{+})\end{equation}
 \begin{equation}
S(y)=\tilde{s}(y)+\sqrt{2}\theta s(y)+\theta^{2}F_{s}(y),~~~S^{+}(y^{+})=\tilde{s}^{*}(y^{+})+\sqrt{2}\bar{\theta}\bar{s}(y^{+})+\bar{\theta}^{2}F_{s}^{*}(y^{+})\end{equation}
 with $y^{m}=x^{m}+i\theta\sigma^{m}\bar{\theta}$ and $y^{+m}=x^{m}-i\theta\sigma^{m}\bar{\theta}$.

Products of chiral superfields are again chiral superfields, and likewise
for their conjugates. Define again \begin{equation}
A(y)=\Phi(y)S(y)=\tilde{a}(y)+\sqrt{2}\theta a(y)+\theta^{2}F_{a}(y)\end{equation}
 \begin{equation}
A^{+}(y^{+})=\Phi^{+}(y^{+})S^{+}(y^{+})=\tilde{a}^{*}(y^{+})+\sqrt{2}\bar{\theta}\bar{a}(y^{+})+\bar{\theta}^{2}F_{a}^{*}(y^{+})\end{equation}
The component fields of these composite superfields are
\begin{equation}
\tilde{a}=\tilde{\psi}\tilde{s},~~a=\tilde{\psi}s+\tilde{s}\psi,~~F_{a}=\tilde{\psi}F_{s}+\tilde{s}F_{\psi}-\psi s\label{eq:new from old1}\end{equation}
 \begin{equation}
\tilde{a}^{*}=\tilde{\psi}^{*}\tilde{s}^{*},~~\bar{a}=\tilde{\psi}^{*}\bar{s}+\tilde{s}^{*}\bar{\psi},~~F_{a}^{*}=\tilde{\psi}^{*}F_{s}^{*}+\tilde{s}^{*}F_{\psi}^{*}-\bar{\psi}\bar{s}\label{eq:new from old2}\end{equation}

The $\theta\theta\bar{\theta}\bar{\theta}$ term of the dimension
six operator \begin{equation}
\sum\frac{S^{+}S\Phi^{+}\Phi}{M_{GUT}^{2}}=\sum\frac{S^{+}\Phi^{+}\Phi S}{M_{GUT}^{2}}=\sum\frac{(\Phi S)^{+}(\Phi S)}{M_{GUT}^{2}}=\sum\frac{A^{+}A}{M_{GUT}^{2}}\end{equation}
 is given as \cite{Wess:1992cp}: \begin{equation}
F_{a}^{*}F_{a}+\frac{1}{4}\tilde{a}^{*}\square\tilde{a}+\frac{1}{4}\square\tilde{a}^{*}\tilde{a}-\frac{1}{2}\partial_{m}\tilde{a}^{*}
\partial^{m}\tilde{a}+\frac{i}{2}\partial_{m}\bar{a}\bar{\sigma}^{m}a-\frac{i}{2}\bar{a}\bar{\sigma}^{m}\partial_{m}a \label{a8} \end{equation}

Assuming the singlet scalar develops a vacuum expectation value (VEV)
$<\widetilde{s}>$, the $Z_{2}$ symmetry is spontaneously broken
and both components $(\widetilde{s},s)$ in $S$ will decay. Eq. (\ref{a8}) can be reexpressed via Eq.(\ref{eq:new from old1},\ref{eq:new from old2}).
\begin{equation}
\underset{\Phi}{\sum}\frac{1}{M_{GUT}^{2}}\left(i<\widetilde{s}>\widetilde{s}^{*}(\partial_{\mu}\psi\sigma^{\mu}\bar{\psi})+i<\widetilde{s}>\widetilde{\psi}^{*}(\partial_{\mu}\psi\sigma^{\mu}\bar{s})+<\widetilde{s}>\widetilde{s}^{*}\widetilde{\psi}^{*}\square\widetilde{\psi}\right)+h.c.+...\end{equation}
 Here we have dropped total divergence terms. Operators from F-terms
have also been neglected as they are suppressed by the leptonic Yukawa
coupling constant. In addition, these operators will lead to many
body decays which are further suppressed by phase spaces.

\bibliographystyle{elsart-num} \bibliographystyle{elsart-num}
\bibliography{fermi}

\begin{thebibliography}{52}
\expandafter\ifx\csname natexlab\endcsname\relax\def\natexlab#1{#1}\fi
\expandafter\ifx\csname bibnamefont\endcsname\relax
  \def\bibnamefont#1{#1}\fi
\expandafter\ifx\csname bibfnamefont\endcsname\relax
  \def\bibfnamefont#1{#1}\fi
\expandafter\ifx\csname citenamefont\endcsname\relax
  \def\citenamefont#1{#1}\fi
\expandafter\ifx\csname url\endcsname\relax
  \def\url#1{\texttt{#1}}\fi
\expandafter\ifx\csname urlprefix\endcsname\relax\def\urlprefix{URL }\fi
\providecommand{\bibinfo}[2]{#2}
\providecommand{\eprint}[2][]{\url{#2}}

\bibitem[{\citenamefont{Adriani et~al.}(2009{\natexlab{a}})}]{Adriani:2008zr}
\bibinfo{author}{\bibfnamefont{O.}~\bibnamefont{Adriani}} \bibnamefont{et~al.}
  (\bibinfo{collaboration}{PAMELA}), \bibinfo{journal}{Nature}
  \textbf{\bibinfo{volume}{458}}, \bibinfo{pages}{607}
  (\bibinfo{year}{2009}{\natexlab{a}}), \eprint{0810.4995}.

\bibitem[{\citenamefont{Adriani et~al.}(2009{\natexlab{b}})}]{Adriani:2008zq}
\bibinfo{author}{\bibfnamefont{O.}~\bibnamefont{Adriani}} \bibnamefont{et~al.},
  \bibinfo{journal}{Phys. Rev. Lett.} \textbf{\bibinfo{volume}{102}},
  \bibinfo{pages}{051101} (\bibinfo{year}{2009}{\natexlab{b}}),
  \eprint{0810.4994}.

\bibitem[{\citenamefont{Abdo et~al.}(2009)}]{Abdo:2009zk}
\bibinfo{author}{\bibfnamefont{A.~A.} \bibnamefont{Abdo}} \bibnamefont{et~al.}
  (\bibinfo{collaboration}{The Fermi LAT}), \bibinfo{journal}{Phys. Rev. Lett.}
  \textbf{\bibinfo{volume}{102}}, \bibinfo{pages}{181101}
  (\bibinfo{year}{2009}), \eprint{0905.0025}.

\bibitem[{\citenamefont{Aharonian}(2009)}]{Aharonian:2009ah}
\bibinfo{author}{\bibfnamefont{H.~E. S. S. C.~F.} \bibnamefont{Aharonian}}
  (\bibinfo{year}{2009}), \eprint{0905.0105}.

\bibitem[{\citenamefont{Stawarz et~al.}(2009)\citenamefont{Stawarz, Petrosian,
  and Blandford}}]{Stawarz:2009ig}
\bibinfo{author}{\bibfnamefont{L.}~\bibnamefont{Stawarz}},
  \bibinfo{author}{\bibfnamefont{V.}~\bibnamefont{Petrosian}},
  \bibnamefont{and} \bibinfo{author}{\bibfnamefont{R.~D.}
  \bibnamefont{Blandford}} (\bibinfo{year}{2009}), \eprint{0908.1094}.

\bibitem[{\citenamefont{Piran et~al.}(2009)\citenamefont{Piran, Shaviv, and
  Nakar}}]{Piran:2009tx}
\bibinfo{author}{\bibfnamefont{T.}~\bibnamefont{Piran}},
  \bibinfo{author}{\bibfnamefont{N.~J.} \bibnamefont{Shaviv}},
  \bibnamefont{and} \bibinfo{author}{\bibfnamefont{E.}~\bibnamefont{Nakar}}
  (\bibinfo{year}{2009}), \eprint{0905.0904}.

\bibitem[{\citenamefont{Grasso et~al.}(2009)}]{Grasso:2009ma}
\bibinfo{author}{\bibfnamefont{D.}~\bibnamefont{Grasso}} \bibnamefont{et~al.}
  (\bibinfo{collaboration}{FERMI-LAT}), \bibinfo{journal}{Astropart. Phys.}
  \textbf{\bibinfo{volume}{32}}, \bibinfo{pages}{140} (\bibinfo{year}{2009}),
  \eprint{0905.0636}.

\bibitem[{\citenamefont{Malyshev et~al.}(2009)\citenamefont{Malyshev, Cholis,
  and Gelfand}}]{Malyshev:2009tw}
\bibinfo{author}{\bibfnamefont{D.}~\bibnamefont{Malyshev}},
  \bibinfo{author}{\bibfnamefont{I.}~\bibnamefont{Cholis}}, \bibnamefont{and}
  \bibinfo{author}{\bibfnamefont{J.}~\bibnamefont{Gelfand}},
  \bibinfo{journal}{Phys. Rev.} \textbf{\bibinfo{volume}{D80}},
  \bibinfo{pages}{063005} (\bibinfo{year}{2009}), \eprint{0903.1310}.

\bibitem[{\citenamefont{Bertone et~al.}(2009)\citenamefont{Bertone, Cirelli,
  Strumia, and Taoso}}]{Bertone:2008xr}
\bibinfo{author}{\bibfnamefont{G.}~\bibnamefont{Bertone}},
  \bibinfo{author}{\bibfnamefont{M.}~\bibnamefont{Cirelli}},
  \bibinfo{author}{\bibfnamefont{A.}~\bibnamefont{Strumia}}, \bibnamefont{and}
  \bibinfo{author}{\bibfnamefont{M.}~\bibnamefont{Taoso}},
  \bibinfo{journal}{JCAP} \textbf{\bibinfo{volume}{0903}}, \bibinfo{pages}{009}
  (\bibinfo{year}{2009}), \eprint{0811.3744}.

\bibitem[{\citenamefont{Zhang et~al.}(2009{\natexlab{a}})}]{Zhang:2008tb}
\bibinfo{author}{\bibfnamefont{J.}~\bibnamefont{Zhang}} \bibnamefont{et~al.},
  \bibinfo{journal}{Phys. Rev.} \textbf{\bibinfo{volume}{D80}},
  \bibinfo{pages}{023007} (\bibinfo{year}{2009}{\natexlab{a}}),
  \eprint{0812.0522}.

\bibitem[{\citenamefont{Bergstrom
  et~al.}(2009{\natexlab{a}})\citenamefont{Bergstrom, Bertone, Bringmann,
  Edsjo, and Taoso}}]{Bergstrom:2008ag}
\bibinfo{author}{\bibfnamefont{L.}~\bibnamefont{Bergstrom}},
  \bibinfo{author}{\bibfnamefont{G.}~\bibnamefont{Bertone}},
  \bibinfo{author}{\bibfnamefont{T.}~\bibnamefont{Bringmann}},
  \bibinfo{author}{\bibfnamefont{J.}~\bibnamefont{Edsjo}}, \bibnamefont{and}
  \bibinfo{author}{\bibfnamefont{M.}~\bibnamefont{Taoso}},
  \bibinfo{journal}{Phys. Rev.} \textbf{\bibinfo{volume}{D79}},
  \bibinfo{pages}{081303} (\bibinfo{year}{2009}{\natexlab{a}}),
  \eprint{0812.3895}.

\bibitem[{\citenamefont{Ibarra et~al.}(2009)\citenamefont{Ibarra, Tran, and
  Weniger}}]{Ibarra:2009nw}
\bibinfo{author}{\bibfnamefont{A.}~\bibnamefont{Ibarra}},
  \bibinfo{author}{\bibfnamefont{D.}~\bibnamefont{Tran}}, \bibnamefont{and}
  \bibinfo{author}{\bibfnamefont{C.}~\bibnamefont{Weniger}}
  (\bibinfo{year}{2009}), \eprint{0909.3514}.

\bibitem[{\citenamefont{Zhang et~al.}(2009{\natexlab{b}})\citenamefont{Zhang,
  Yuan, and Bi}}]{Zhang:2009kp}
\bibinfo{author}{\bibfnamefont{J.}~\bibnamefont{Zhang}},
  \bibinfo{author}{\bibfnamefont{Q.}~\bibnamefont{Yuan}}, \bibnamefont{and}
  \bibinfo{author}{\bibfnamefont{X.-J.} \bibnamefont{Bi}}
  (\bibinfo{year}{2009}{\natexlab{b}}), \eprint{0908.1236}.

\bibitem[{\citenamefont{Meade et~al.}(2009)\citenamefont{Meade, Papucci,
  Strumia, and Volansky}}]{Meade:2009iu}
\bibinfo{author}{\bibfnamefont{P.}~\bibnamefont{Meade}},
  \bibinfo{author}{\bibfnamefont{M.}~\bibnamefont{Papucci}},
  \bibinfo{author}{\bibfnamefont{A.}~\bibnamefont{Strumia}}, \bibnamefont{and}
  \bibinfo{author}{\bibfnamefont{T.}~\bibnamefont{Volansky}}
  (\bibinfo{year}{2009}), \eprint{0905.0480}.

\bibitem[{\citenamefont{Bergstrom
  et~al.}(2009{\natexlab{b}})\citenamefont{Bergstrom, Edsjo, and
  Zaharijas}}]{Bergstrom:2009fa}
\bibinfo{author}{\bibfnamefont{L.}~\bibnamefont{Bergstrom}},
  \bibinfo{author}{\bibfnamefont{J.}~\bibnamefont{Edsjo}}, \bibnamefont{and}
  \bibinfo{author}{\bibfnamefont{G.}~\bibnamefont{Zaharijas}},
  \bibinfo{journal}{Phys. Rev. Lett.} \textbf{\bibinfo{volume}{103}},
  \bibinfo{pages}{031103} (\bibinfo{year}{2009}{\natexlab{b}}),
  \eprint{0905.0333}.

\bibitem[{\citenamefont{Dunkley et~al.}(2009)}]{Dunkley:2008ie}
\bibinfo{author}{\bibfnamefont{J.}~\bibnamefont{Dunkley}} \bibnamefont{et~al.}
  (\bibinfo{collaboration}{WMAP}), \bibinfo{journal}{Astrophys. J. Suppl.}
  \textbf{\bibinfo{volume}{180}}, \bibinfo{pages}{306} (\bibinfo{year}{2009}),
  \eprint{0803.0586}.

\bibitem[{\citenamefont{Hisano et~al.}(2005)\citenamefont{Hisano, Matsumoto,
  Nojiri, and Saito}}]{Hisano:2004ds}
\bibinfo{author}{\bibfnamefont{J.}~\bibnamefont{Hisano}},
  \bibinfo{author}{\bibfnamefont{S.}~\bibnamefont{Matsumoto}},
  \bibinfo{author}{\bibfnamefont{M.~M.} \bibnamefont{Nojiri}},
  \bibnamefont{and} \bibinfo{author}{\bibfnamefont{O.}~\bibnamefont{Saito}},
  \bibinfo{journal}{Phys. Rev.} \textbf{\bibinfo{volume}{D71}},
  \bibinfo{pages}{063528} (\bibinfo{year}{2005}), \eprint{hep-ph/0412403}.

\bibitem[{\citenamefont{Hisano et~al.}(2007)\citenamefont{Hisano, Matsumoto,
  Nagai, Saito, and Senami}}]{Hisano:2006nn}
\bibinfo{author}{\bibfnamefont{J.}~\bibnamefont{Hisano}},
  \bibinfo{author}{\bibfnamefont{S.}~\bibnamefont{Matsumoto}},
  \bibinfo{author}{\bibfnamefont{M.}~\bibnamefont{Nagai}},
  \bibinfo{author}{\bibfnamefont{O.}~\bibnamefont{Saito}}, \bibnamefont{and}
  \bibinfo{author}{\bibfnamefont{M.}~\bibnamefont{Senami}},
  \bibinfo{journal}{Phys. Lett.} \textbf{\bibinfo{volume}{B646}},
  \bibinfo{pages}{34} (\bibinfo{year}{2007}), \eprint{hep-ph/0610249}.

\bibitem[{\citenamefont{Cirelli et~al.}(2007)\citenamefont{Cirelli, Strumia,
  and Tamburini}}]{Cirelli:2007xd}
\bibinfo{author}{\bibfnamefont{M.}~\bibnamefont{Cirelli}},
  \bibinfo{author}{\bibfnamefont{A.}~\bibnamefont{Strumia}}, \bibnamefont{and}
  \bibinfo{author}{\bibfnamefont{M.}~\bibnamefont{Tamburini}},
  \bibinfo{journal}{Nucl. Phys.} \textbf{\bibinfo{volume}{B787}},
  \bibinfo{pages}{152} (\bibinfo{year}{2007}), \eprint{0706.4071}.

\bibitem[{\citenamefont{Arkani-Hamed et~al.}(2009)\citenamefont{Arkani-Hamed,
  Finkbeiner, Slatyer, and Weiner}}]{ArkaniHamed:2008qn}
\bibinfo{author}{\bibfnamefont{N.}~\bibnamefont{Arkani-Hamed}},
  \bibinfo{author}{\bibfnamefont{D.~P.} \bibnamefont{Finkbeiner}},
  \bibinfo{author}{\bibfnamefont{T.~R.} \bibnamefont{Slatyer}},
  \bibnamefont{and} \bibinfo{author}{\bibfnamefont{N.}~\bibnamefont{Weiner}},
  \bibinfo{journal}{Phys. Rev.} \textbf{\bibinfo{volume}{D79}},
  \bibinfo{pages}{015014} (\bibinfo{year}{2009}), \eprint{0810.0713}.

\bibitem[{\citenamefont{Feldman et~al.}(2009)\citenamefont{Feldman, Liu, and
  Nath}}]{Feldman:2008xs}
\bibinfo{author}{\bibfnamefont{D.}~\bibnamefont{Feldman}},
  \bibinfo{author}{\bibfnamefont{Z.}~\bibnamefont{Liu}}, \bibnamefont{and}
  \bibinfo{author}{\bibfnamefont{P.}~\bibnamefont{Nath}},
  \bibinfo{journal}{Phys. Rev.} \textbf{\bibinfo{volume}{D79}},
  \bibinfo{pages}{063509} (\bibinfo{year}{2009}), \eprint{0810.5762}.

\bibitem[{\citenamefont{Ibe et~al.}(2009)\citenamefont{Ibe, Murayama, and
  Yanagida}}]{Ibe:2008ye}
\bibinfo{author}{\bibfnamefont{M.}~\bibnamefont{Ibe}},
  \bibinfo{author}{\bibfnamefont{H.}~\bibnamefont{Murayama}}, \bibnamefont{and}
  \bibinfo{author}{\bibfnamefont{T.~T.} \bibnamefont{Yanagida}},
  \bibinfo{journal}{Phys. Rev.} \textbf{\bibinfo{volume}{D79}},
  \bibinfo{pages}{095009} (\bibinfo{year}{2009}), \eprint{0812.0072}.

\bibitem[{\citenamefont{Guo and Wu}(2009)}]{Guo:2009aj}
\bibinfo{author}{\bibfnamefont{W.-L.} \bibnamefont{Guo}} \bibnamefont{and}
  \bibinfo{author}{\bibfnamefont{Y.-L.} \bibnamefont{Wu}},
  \bibinfo{journal}{Phys. Rev.} \textbf{\bibinfo{volume}{D79}},
  \bibinfo{pages}{055012} (\bibinfo{year}{2009}), \eprint{0901.1450}.

\bibitem[{\citenamefont{Shirai et~al.}(2009)\citenamefont{Shirai, Takahashi,
  and Yanagida}}]{Shirai:2009fq}
\bibinfo{author}{\bibfnamefont{S.}~\bibnamefont{Shirai}},
  \bibinfo{author}{\bibfnamefont{F.}~\bibnamefont{Takahashi}},
  \bibnamefont{and} \bibinfo{author}{\bibfnamefont{T.~T.}
  \bibnamefont{Yanagida}}, \bibinfo{journal}{Phys. Lett.}
  \textbf{\bibinfo{volume}{B680}}, \bibinfo{pages}{485} (\bibinfo{year}{2009}),
  \eprint{0905.0388}.

\bibitem[{\citenamefont{Mardon et~al.}(2009)\citenamefont{Mardon, Nomura, and
  Thaler}}]{Mardon:2009gw}
\bibinfo{author}{\bibfnamefont{J.}~\bibnamefont{Mardon}},
  \bibinfo{author}{\bibfnamefont{Y.}~\bibnamefont{Nomura}}, \bibnamefont{and}
  \bibinfo{author}{\bibfnamefont{J.}~\bibnamefont{Thaler}},
  \bibinfo{journal}{Phys. Rev.} \textbf{\bibinfo{volume}{D80}},
  \bibinfo{pages}{035013} (\bibinfo{year}{2009}), \eprint{0905.3749}.

\bibitem[{\citenamefont{Langacker and Luo}(1991)}]{Langacker:1991an}
\bibinfo{author}{\bibfnamefont{P.}~\bibnamefont{Langacker}} \bibnamefont{and}
  \bibinfo{author}{\bibfnamefont{M.-x.} \bibnamefont{Luo}},
  \bibinfo{journal}{Phys. Rev.} \textbf{\bibinfo{volume}{D44}},
  \bibinfo{pages}{817} (\bibinfo{year}{1991}).

\bibitem[{\citenamefont{Amaldi et~al.}(1991)\citenamefont{Amaldi, de~Boer, and
  Furstenau}}]{Amaldi:1991cn}
\bibinfo{author}{\bibfnamefont{U.}~\bibnamefont{Amaldi}},
  \bibinfo{author}{\bibfnamefont{W.}~\bibnamefont{de~Boer}}, \bibnamefont{and}
  \bibinfo{author}{\bibfnamefont{H.}~\bibnamefont{Furstenau}},
  \bibinfo{journal}{Phys. Lett.} \textbf{\bibinfo{volume}{B260}},
  \bibinfo{pages}{447} (\bibinfo{year}{1991}).

\bibitem[{\citenamefont{Ellis et~al.}(1991)\citenamefont{Ellis, Kelley, and
  Nanopoulos}}]{Ellis:1990wk}
\bibinfo{author}{\bibfnamefont{J.~R.} \bibnamefont{Ellis}},
  \bibinfo{author}{\bibfnamefont{S.}~\bibnamefont{Kelley}}, \bibnamefont{and}
  \bibinfo{author}{\bibfnamefont{D.~V.} \bibnamefont{Nanopoulos}},
  \bibinfo{journal}{Phys. Lett.} \textbf{\bibinfo{volume}{B260}},
  \bibinfo{pages}{131} (\bibinfo{year}{1991}).

\bibitem[{\citenamefont{Arvanitaki
  et~al.}(2009{\natexlab{a}})}]{Arvanitaki:2008hq}
\bibinfo{author}{\bibfnamefont{A.}~\bibnamefont{Arvanitaki}}
  \bibnamefont{et~al.}, \bibinfo{journal}{Phys. Rev.}
  \textbf{\bibinfo{volume}{D79}}, \bibinfo{pages}{105022}
  (\bibinfo{year}{2009}{\natexlab{a}}), \eprint{0812.2075}.

\bibitem[{\citenamefont{Arvanitaki
  et~al.}(2009{\natexlab{b}})}]{Arvanitaki:2009yb}
\bibinfo{author}{\bibfnamefont{A.}~\bibnamefont{Arvanitaki}}
  \bibnamefont{et~al.}, \bibinfo{journal}{Phys. Rev.}
  \textbf{\bibinfo{volume}{D80}}, \bibinfo{pages}{055011}
  (\bibinfo{year}{2009}{\natexlab{b}}), \eprint{0904.2789}.

\bibitem[{\citenamefont{Porter and Collaboration}(2009)}]{Porter:2009sg}
\bibinfo{author}{\bibfnamefont{T.~A.} \bibnamefont{Porter}} \bibnamefont{and}
  \bibinfo{author}{\bibfnamefont{f.~t. F.~L.} \bibnamefont{Collaboration}}
  (\bibinfo{year}{2009}), \eprint{0907.0294}.

\bibitem[{\citenamefont{Navarro et~al.}(1997)\citenamefont{Navarro, Frenk, and
  White}}]{Navarro:1996gj}
\bibinfo{author}{\bibfnamefont{J.~F.} \bibnamefont{Navarro}},
  \bibinfo{author}{\bibfnamefont{C.~S.} \bibnamefont{Frenk}}, \bibnamefont{and}
  \bibinfo{author}{\bibfnamefont{S.~D.~M.} \bibnamefont{White}},
  \bibinfo{journal}{Astrophys. J.} \textbf{\bibinfo{volume}{490}},
  \bibinfo{pages}{493} (\bibinfo{year}{1997}), \eprint{astro-ph/9611107}.

\bibitem[{\citenamefont{Donato et~al.}(2004)\citenamefont{Donato, Fornengo,
  Maurin, and Salati}}]{Donato:2003xg}
\bibinfo{author}{\bibfnamefont{F.}~\bibnamefont{Donato}},
  \bibinfo{author}{\bibfnamefont{N.}~\bibnamefont{Fornengo}},
  \bibinfo{author}{\bibfnamefont{D.}~\bibnamefont{Maurin}}, \bibnamefont{and}
  \bibinfo{author}{\bibfnamefont{P.}~\bibnamefont{Salati}},
  \bibinfo{journal}{Phys. Rev.} \textbf{\bibinfo{volume}{D69}},
  \bibinfo{pages}{063501} (\bibinfo{year}{2004}), \eprint{astro-ph/0306207}.

\bibitem[{\citenamefont{Delahaye et~al.}(2008)\citenamefont{Delahaye, Lineros,
  Donato, Fornengo, and Salati}}]{Delahaye:2007fr}
\bibinfo{author}{\bibfnamefont{T.}~\bibnamefont{Delahaye}},
  \bibinfo{author}{\bibfnamefont{R.}~\bibnamefont{Lineros}},
  \bibinfo{author}{\bibfnamefont{F.}~\bibnamefont{Donato}},
  \bibinfo{author}{\bibfnamefont{N.}~\bibnamefont{Fornengo}}, \bibnamefont{and}
  \bibinfo{author}{\bibfnamefont{P.}~\bibnamefont{Salati}},
  \bibinfo{journal}{Phys. Rev.} \textbf{\bibinfo{volume}{D77}},
  \bibinfo{pages}{063527} (\bibinfo{year}{2008}), \eprint{0712.2312}.

\bibitem[{\citenamefont{Ruderman and
  Volansky}(2009{\natexlab{a}})}]{Ruderman:2009ta}
\bibinfo{author}{\bibfnamefont{J.~T.} \bibnamefont{Ruderman}} \bibnamefont{and}
  \bibinfo{author}{\bibfnamefont{T.}~\bibnamefont{Volansky}}
  (\bibinfo{year}{2009}{\natexlab{a}}), \eprint{0907.4373}.

\bibitem[{\citenamefont{Ruderman and
  Volansky}(2009{\natexlab{b}})}]{Ruderman:2009tj}
\bibinfo{author}{\bibfnamefont{J.~T.} \bibnamefont{Ruderman}} \bibnamefont{and}
  \bibinfo{author}{\bibfnamefont{T.}~\bibnamefont{Volansky}}
  (\bibinfo{year}{2009}{\natexlab{b}}), \eprint{0908.1570}.

\bibitem[{\citenamefont{Kadastik et~al.}(2009)\citenamefont{Kadastik, Kannike,
  and Raidal}}]{Kadastik:2009cu}
\bibinfo{author}{\bibfnamefont{M.}~\bibnamefont{Kadastik}},
  \bibinfo{author}{\bibfnamefont{K.}~\bibnamefont{Kannike}}, \bibnamefont{and}
  \bibinfo{author}{\bibfnamefont{M.}~\bibnamefont{Raidal}}
  (\bibinfo{year}{2009}), \eprint{0907.1894}.

\bibitem[{\citenamefont{Kyae}(2009)}]{Kyae:2009gm}
\bibinfo{author}{\bibfnamefont{B.}~\bibnamefont{Kyae}} (\bibinfo{year}{2009}),
  \eprint{0909.3139}.

\bibitem[{\citenamefont{Sjostrand et~al.}(2006)\citenamefont{Sjostrand, Mrenna,
  and Skands}}]{Sjostrand:2006za}
\bibinfo{author}{\bibfnamefont{T.}~\bibnamefont{Sjostrand}},
  \bibinfo{author}{\bibfnamefont{S.}~\bibnamefont{Mrenna}}, \bibnamefont{and}
  \bibinfo{author}{\bibfnamefont{P.}~\bibnamefont{Skands}},
  \bibinfo{journal}{JHEP} \textbf{\bibinfo{volume}{05}}, \bibinfo{pages}{026}
  (\bibinfo{year}{2006}), \eprint{hep-ph/0603175}.

\bibitem[{\citenamefont{Cirelli et~al.}(2008)\citenamefont{Cirelli,
  Franceschini, and Strumia}}]{Cirelli:2008id}
\bibinfo{author}{\bibfnamefont{M.}~\bibnamefont{Cirelli}},
  \bibinfo{author}{\bibfnamefont{R.}~\bibnamefont{Franceschini}},
  \bibnamefont{and} \bibinfo{author}{\bibfnamefont{A.}~\bibnamefont{Strumia}},
  \bibinfo{journal}{Nucl. Phys.} \textbf{\bibinfo{volume}{B800}},
  \bibinfo{pages}{204} (\bibinfo{year}{2008}), \eprint{0802.3378}.

\bibitem[{\citenamefont{Cheung et~al.}(2009)\citenamefont{Cheung, Tseng, and
  Yuan}}]{Cheung:2009si}
\bibinfo{author}{\bibfnamefont{K.}~\bibnamefont{Cheung}},
  \bibinfo{author}{\bibfnamefont{P.-Y.} \bibnamefont{Tseng}}, \bibnamefont{and}
  \bibinfo{author}{\bibfnamefont{T.-C.} \bibnamefont{Yuan}},
  \bibinfo{journal}{Phys. Lett.} \textbf{\bibinfo{volume}{B678}},
  \bibinfo{pages}{293} (\bibinfo{year}{2009}), \eprint{0902.4035}.

\bibitem[{\citenamefont{Ishiwata
  et~al.}(2009{\natexlab{a}})\citenamefont{Ishiwata, Matsumoto, and
  Moroi}}]{Ishiwata:2009vx}
\bibinfo{author}{\bibfnamefont{K.}~\bibnamefont{Ishiwata}},
  \bibinfo{author}{\bibfnamefont{S.}~\bibnamefont{Matsumoto}},
  \bibnamefont{and} \bibinfo{author}{\bibfnamefont{T.}~\bibnamefont{Moroi}},
  \bibinfo{journal}{JHEP} \textbf{\bibinfo{volume}{05}}, \bibinfo{pages}{110}
  (\bibinfo{year}{2009}{\natexlab{a}}), \eprint{0903.0242}.

\bibitem[{\citenamefont{Ibarra et~al.}(2010)\citenamefont{Ibarra, Tran, and
  Weniger}}]{Ibarra:2009dr}
\bibinfo{author}{\bibfnamefont{A.}~\bibnamefont{Ibarra}},
  \bibinfo{author}{\bibfnamefont{D.}~\bibnamefont{Tran}}, \bibnamefont{and}
  \bibinfo{author}{\bibfnamefont{C.}~\bibnamefont{Weniger}},
  \bibinfo{journal}{JCAP} \textbf{\bibinfo{volume}{1001}}, \bibinfo{pages}{009}
  (\bibinfo{year}{2010}), \eprint{0906.1571}.

\bibitem[{\citenamefont{Perko}(1987)}]{Perko:1987pq}
\bibinfo{author}{\bibfnamefont{J.~S.} \bibnamefont{Perko}},
  \bibinfo{journal}{Astron. Astrophys.} \textbf{\bibinfo{volume}{184}},
  \bibinfo{pages}{119} (\bibinfo{year}{1987}).

\bibitem[{\citenamefont{Gondolo et~al.}(2004)}]{Gondolo:2004sc}
\bibinfo{author}{\bibfnamefont{P.}~\bibnamefont{Gondolo}} \bibnamefont{et~al.},
  \bibinfo{journal}{JCAP} \textbf{\bibinfo{volume}{0407}}, \bibinfo{pages}{008}
  (\bibinfo{year}{2004}), \eprint{astro-ph/0406204}.

\bibitem[{\citenamefont{Blumenthal and Gould}(1970)}]{Blumenthal:1970gc}
\bibinfo{author}{\bibfnamefont{G.~R.} \bibnamefont{Blumenthal}}
  \bibnamefont{and} \bibinfo{author}{\bibfnamefont{R.~J.} \bibnamefont{Gould}},
  \bibinfo{journal}{Rev. Mod. Phys.} \textbf{\bibinfo{volume}{42}},
  \bibinfo{pages}{237} (\bibinfo{year}{1970}).

\bibitem[{\citenamefont{Cirelli and Panci}(2009)}]{Cirelli:2009vg}
\bibinfo{author}{\bibfnamefont{M.}~\bibnamefont{Cirelli}} \bibnamefont{and}
  \bibinfo{author}{\bibfnamefont{P.}~\bibnamefont{Panci}},
  \bibinfo{journal}{Nucl. Phys.} \textbf{\bibinfo{volume}{B821}},
  \bibinfo{pages}{399} (\bibinfo{year}{2009}), \eprint{0904.3830}.

\bibitem[{\citenamefont{Ishiwata
  et~al.}(2009{\natexlab{b}})\citenamefont{Ishiwata, Matsumoto, and
  Moroi}}]{Ishiwata:2009dk}
\bibinfo{author}{\bibfnamefont{K.}~\bibnamefont{Ishiwata}},
  \bibinfo{author}{\bibfnamefont{S.}~\bibnamefont{Matsumoto}},
  \bibnamefont{and} \bibinfo{author}{\bibfnamefont{T.}~\bibnamefont{Moroi}},
  \bibinfo{journal}{Phys. Lett.} \textbf{\bibinfo{volume}{B679}},
  \bibinfo{pages}{1} (\bibinfo{year}{2009}{\natexlab{b}}), \eprint{0905.4593}.

\bibitem[{\citenamefont{Chen et~al.}(2010)\citenamefont{Chen, Mandal, and
  Takahashi}}]{Chen:2009uq}
\bibinfo{author}{\bibfnamefont{C.-R.} \bibnamefont{Chen}},
  \bibinfo{author}{\bibfnamefont{S.~K.} \bibnamefont{Mandal}},
  \bibnamefont{and}
  \bibinfo{author}{\bibfnamefont{F.}~\bibnamefont{Takahashi}},
  \bibinfo{journal}{JCAP} \textbf{\bibinfo{volume}{1001}}, \bibinfo{pages}{023}
  (\bibinfo{year}{2010}), \eprint{0910.2639}.

\bibitem[{\citenamefont{Stecker et~al.}(2006)\citenamefont{Stecker, Malkan, and
  Scully}}]{Stecker:2005qs}
\bibinfo{author}{\bibfnamefont{F.~W.} \bibnamefont{Stecker}},
  \bibinfo{author}{\bibfnamefont{M.~A.} \bibnamefont{Malkan}},
  \bibnamefont{and} \bibinfo{author}{\bibfnamefont{S.~T.}
  \bibnamefont{Scully}}, \bibinfo{journal}{Astrophys. J.}
  \textbf{\bibinfo{volume}{648}}, \bibinfo{pages}{774} (\bibinfo{year}{2006}),
  \eprint{astro-ph/0510449}.

\bibitem[{\citenamefont{Stecker et~al.}(2007)\citenamefont{Stecker, Malkan, and
  Scully}}]{Stecker:2006eh}
\bibinfo{author}{\bibfnamefont{F.~W.} \bibnamefont{Stecker}},
  \bibinfo{author}{\bibfnamefont{M.~A.} \bibnamefont{Malkan}},
  \bibnamefont{and} \bibinfo{author}{\bibfnamefont{S.~T.}
  \bibnamefont{Scully}}, \bibinfo{journal}{Astrophys. J.}
  \textbf{\bibinfo{volume}{658}}, \bibinfo{pages}{1392} (\bibinfo{year}{2007}),
  \eprint{astro-ph/0612048}.

\bibitem[{\citenamefont{Wess and Bagger}(1992)}]{Wess:1992cp}
\bibinfo{author}{\bibfnamefont{J.}~\bibnamefont{Wess}} \bibnamefont{and}
  \bibinfo{author}{\bibfnamefont{J.}~\bibnamefont{Bagger}},
  \bibinfo{journal}{Supersymmetry and supergravity, Princeton University Press}  (\bibinfo{year}{1992}).

\end{thebibliography}

\end{document}